\documentclass[twocolumn,showpacs,pre,showkeys,nofootinbib,a4paper]{revtex4}

\usepackage{graphicx}
\usepackage{latexsym}
\usepackage{amsmath}
\usepackage{amsfonts}

\graphicspath{{./figs/}}

\providecommand{\EE}[1]{\ensuremath{\,{\times}\,10^{#1}}} 
\newcommand{\dig}{\phantom{0}}  

\newcommand{\Pm}{{P}_{\rm m}}

\newcommand{\vect}[1]{{\bf#1}}

\newcommand{\Av}       {\vect{A}}

\newcommand{\Bhatphi}  {\widehat{B}_\varphi}
\newcommand{\Bhatrho}  {\widehat{B}_\RHO}
\newcommand{\Bhatz}    {\widehat{B}_z}

\newcommand{\Bphi}     {B_\varphi}
\newcommand{\Brho}     {B_\RHO}

\newcommand{\Bv}       {\vect{B}}
\newcommand{\Bz}       {B_z}

\newcommand{\csound}   {c_{\rm s}}
\newcommand{\curl}     {\nabla\times}

\newcommand{\Div}      {\nabla\cdot}
\newcommand{\Dop}      {\hat{D}}

\newcommand{\Ev}       {\vect{E}}

\newcommand{\FLor}     {\vect{F}_{\rm L}}

\newcommand{\gammac}   {\lambda}
\newcommand{\gammai}   {\omega}
\newcommand{\gammar}   {\gamma}
\newcommand{\grad}     {\nabla}

\newcommand{\jv}       {\bf j}

\newcommand{\RHO}      {r}

\newcommand{\Reynolds} {{\rm Re}}
\newcommand{\Rm}   {{R}_{\rm m}}
\newcommand{\Rmstar}   {{R}_{\rm m}^{\rm(cr)}}

\newcommand{\Rmtilde}  {\widetilde{R}_{\rm m}}
\newcommand{\rot}      {\curl}

\newcommand{\uv}       {\vect{u}}

\newcommand{\Uz}       {W}

\newcommand{\zerov}    {\vect{0}}

\newcommand{\mbf}{\boldmath}

\newcommand{\dispfrac}[2]{\displaystyle\frac{#1}{#2}}
\newcommand{\ecs}{\extracolsep{1ex}} 

\begin{document}

\preprint{NORDITA-2001-10 AP}

\title{Nonlinear States of the Screw Dynamo}

\author{Wolfgang Dobler}
  \email{Wolfgang.Dobler@kis.uni-freiburg.de}
\author{Anvar Shukurov}
  \email{Anvar.Shukurov@ncl.ac.uk}
\affiliation{Department of Mathematics, University of Newcastle, Newcastle
 upon Tyne, NE1 7RU, U.K.}
\author{Axel Brandenburg}
  \email{brandenb@nordita.dk}
  \altaffiliation[also at: ]{Nordita, Blegdamsvej 17, DK-2100 Copenhagen \O, Denmark}
\affiliation{Department of Mathematics, University of Newcastle, Newcastle
 upon Tyne, NE1 7RU, U.K.}

\date{\today}

\begin{abstract}
  The self-excitation of magnetic field by a spiral Couette flow between
  two coaxial cylinders is considered.
  We solve numerically the fully nonlinear, three-dimensional MHD
  equations for magnetic Prandtl numbers $\Pm$ (ratio of kinematic
  viscosity to magnetic diffusivity) between 0.14 and 10
  and kinematic and magnetic Reynolds numbers up to about 2000.
  In the initial stage of exponential field growth (kinematic dynamo
  regime), we
  find that the dynamo
  switches from one distinct regime to another as the radial width
  $\delta\RHO_B$ of the
  magnetic field distribution becomes smaller than the separation of
  the field maximum from the flow boundary.
  The saturation of magnetic field growth is due to a
  reduction in the velocity shear resulting mainly from
  the azimuthally and axially averaged part of the Lorentz force,
  which agrees with an asymptotic result for the limit
  of $\Pm\ll1$.
  In the parameter regime considered, the magnetic energy decreases with
  kinematic Reynolds number as $\Reynolds^{-0.84}$, which is approximately as
  predicted by the nonlinear asymptotic theory ($\sim\Reynolds^{-1}$).
  However, when the velocity field is maintained by a volume force (rather than
  by viscous stress) the dependence of magnetic energy on the kinematic Reynolds
  number is much weaker.
\end{abstract}



\maketitle

\section{Introduction}

The screw dynamo is a system where magnetic field is
generated by the (laminar) flow of an electrically neutral,
but conducting fluid with helical streamlines, i.e.
\begin{equation}\label{screwflow}
  \uv = ( 0, \RHO\Omega, u_z)
\end{equation}
in cylindrical polar coordinates $(\RHO,\varphi,z)$,
with $\Omega$ and $u_z$ the angular and axial velocities, respectively.
It is one of the simplest dynamo systems known and the most symmetric one
in the sense that the flow can be steady and uniform in
the azimuthal and axial directions.
As first shown by \citet{Lortz:ExactSol,Lortz:SimpleStat}
and \citet{Ponomarenko:Theory}, such a flow can generate magnetic fields
via dynamo action, i.e., without any external electromotive
forces.
Since the magnetic Reynolds number required for magnetic field generation
by the screw dynamo
is relatively low, this type of flow has been used in a series of
laboratory dynamo
experiments in Riga (e.g.\ Refs.~\citep{%
GailitisFreiberg:TheoryScrewDynamo,%
GailitisFreiberg:Nonhomogeneous})
which have recently achieved magnetic field growth and saturation
\citep{GailitisEtal:FieldDetection,GailitisEtal:Saturation}.
There are further plans to perform a dynamo experiment based on
a similar (but time-dependent) flow
\citep{DenisovEtal:FeasabilityR,FrickEtal:Nonstationary}.
Dynamo action of this type
can also occur in the cooling systems of fast
breeder reactors \cite{AMPS2000}.
A related successful dynamo experiment is the Karlsruhe liquid sodium
facility \citep{BusseEtal:Two-Scale,RaedlerEtal:Karlsruhe-Trest}, which
involves an ensemble of spiral flows.
Since the magnetic Reynolds numbers achievable in laboratory flows are
never very high, it is important to understand quantitatively the
excitation properties of the system and to predict measurable
characteristics of the dynamo including the strength, location and time
dependence of the magnetic field in the nonlinear regime.

Other possible sites for screw dynamo action are astrophysical jets
\citep{ShukurovSokoloff:Hydromagnetic} where a helical flow capable of
dynamo action can arise from the axial ejection of plasma from a rotating
accretion disc \citep{KP2000}.

A discussion of the screw dynamo in a broader context of slow dynamos was
presented by \citet{Soward:Unified}.
\citet{GilbertPonty:StreamSurfs} generalized the idea of the screw dynamo
to certain non-axisymmetric flows and
\citet{PontyEtal:ShearConvection} applied this approach
to hydrodynamically unstable Ekman layers.

In the present paper, we explore nonlinear states of the screw dynamo in
the spiral Couette
flow of a viscous fluid between two coaxial cylinders.
Both the screw dynamo itself and the flow are simple enough to allow
detailed analysis of the nonlinear behavior --- a rare feature among MHD
dynamo systems.
In particular, this allows one to assess many of the empirical and
heuristic arguments often applied to more
complicated dynamo systems, such as the relevance of the marginally
stable linear solution for the description of nonlinear states, and to
understand the nonlinear states in considerable detail.

The plan of the paper is as follows.
We briefly review previous studies of the
screw dynamo in Sect.~\ref{Sec-Linear-Theory},
and then describe our model in Sect.~\ref{Sec-Model}.
Our results are presented in
Sect.~\ref{Sec-Linear-Results} for the (kinematic) stage of exponential growth
and in Sect.~\ref{Sec-Nonlin-Results} for saturated, non-linear states.
The results are summarized in Sect.~\ref{Sec-Conclusions}.

\section{The screw dynamo}
\label{Sec-Linear-Theory}

The kinematic behavior of the screw dynamo, including that in the spiral
Couette flow, is
well explored using both asymptotic analysis
\cite{%
  RuzmaikinEtal:HydroScrew,%
  Gilbert:FastPonomarenko,%
  Gilbert:Diss,%
  RuzmaikinEtal:Couette-Poiseuille,%
  SokoloffEtAl:SecondApprox%
}
and numerical mo\-de\-ling
\citep{%
  Solovyov:Description,%
  Solovyov:Excitation2,%
  LupyanShukurov:Screw,%
  Leorat:Modelling%
}.
Consider a time-independent velocity field (\ref{screwflow}) where
both the angular and axial velocity are functions of cylindrical radius
alone, $\Omega=\Omega(\RHO)$ and $u_z=u_z(\RHO)$.
The evolution of the magnetic field
$\Bv$ is governed by the induction equation
\begin{equation}\label{ind}
  \frac{\partial\Bv}{\partial t}
  = \curl\left( \uv\times\Bv -\eta\curl\Bv \right) ,
\end{equation}
supplemented with appropriate boundary conditions.
Here $\eta$ is the magnetic diffusivity, which is related to the
electrical conductivity $\sigma$ of the medium as $\eta = 1/(\mu_0\sigma)$.
At the kinematic stage, when the magnetic field is weak enough,
$\uv$ can be considered fixed and independent of $\Bv$.
The magnetic field can then grow
exponentially provided the magnetic Reynolds number is above a certain
critical value $\Rmstar$, and Eq.~(\ref{ind}) becomes an eigenvalue problem.
The field is necessarily non-axisymmetric (in accordance with
Cowling's theorem, e.g.\ Ref.~\citep{KrauseRaedler:MeanFieldBook})
and, due to the symmetry
of the flow, is a superposition of eigenmodes given in cylindrical
polar coordinates by
\begin{equation} \label{B-harmon}
  B_j(\RHO,\varphi,z,t)
  = \hat{B}_j(\RHO) \, e^{{\rm i}(m\varphi+kz)+\gammac t},\quad
  j={\RHO,\varphi,z} ,
\end{equation}
where $m$ and $k$ are the azimuthal and axial wavenumbers, respectively,
and
\[
  \gammac = \gammar + {\rm i}\gammai
\]
is the eigenvalue, with $\gamma$ the growth rate and $\gammai$ the
oscillation frequency of the magnetic field.

While \citet{Ponomarenko:Theory} discussed a rigid cylinder moving in
a conducting medium --- thus giving rise to a discontinuous
velocity profile --- later models
\cite{RuzmaikinEtal:HydroScrew,Gilbert:FastPonomarenko,Soward:Unified}
apply to more realistic, continuous velocity fields like the
spiral Couette--Poiseuille flow, of which the spiral Couette flow used
in the present paper is a special case.

The coupling of the radial and azimuthal components of Eq.~(\ref{ind}),
required for the magnetic field to grow ($\gamma>0$), occurs via
the diffusion term and is thus proportional to $\eta$
--- see Eqs.~(\ref{Pono-1d-rho}) and (\ref{Pono-1d-phi}).
Therefore, the
growth rate of any given magnetic eigenmode (i.e., for fixed $m$ and $k$)
tends to zero as $\eta\to0$.
The scaling of the growth rate with the magnetic Reynolds number
$\Rm\propto\eta^{-1}$ depends on the flow properties.
In the asymptotic
limit $\Rm\gg1$, $\gamma={\cal O}(\Rm^{-1/2})$ for a continuous
velocity field \cite{RuzmaikinEtal:HydroScrew}, whereas
$\gamma={\cal O}(\Rm^{-1/3})$ for a discontinuous
velocity field \cite{Ponomarenko:Theory,ZRS1983}.
The eigenfunction has a maximum at a radius $\RHO_0$ where the
advection term $m\Omega(\RHO) + k u_z(\RHO)$ [see Eqs.~(\ref{Pono-1d-rho}) and
(\ref{Pono-1d-phi}) in Appendix~A]
has an extremum in $\RHO$, thus minimizing destruction of the magnetic
structure by the $\RHO$-dependent advection.
This implies that $\RHO_0$ satisfies
\begin{equation}
  m\Omega'(\RHO_0) + k u_z'(\RHO_0) = 0 ,
  \label{asymp-r0}
\end{equation}
where primes denote the derivative with respect to $\RHO$. (In a discontinuous
flow, the eigenfunction is localized at the discontinuity.)
Modes with different ratios $k/m$ are localized
at different radii.
An additional necessary condition for the existence of growing modes,
which is due to \cite{Gilbert:FastPonomarenko}, requires that
\begin{equation}
  \left| \frac{{\rm d} \ln\left|\Omega'/u_z'\right|}
                {{\rm d} \ln \RHO}
  \right|
  < 4
\end{equation}
is satisfied at $\RHO=\RHO_0$. This is always the case for the spiral
Couette flow.

The oscillation frequency of a mode localized at $\RHO=\RHO_0$ is
dominated by the advection term,
\[
  \omega=-m\Omega(\RHO_0)-ku_z(\RHO_0)+{\cal O}(\Rm^{-1/2}) ,
\]
for a continuous flow with $m,k={\cal O}(1)$.

The critical magnetic Reynolds number $\Rmstar$,
above which $\gamma>0$, depends on the radial velocity
profile and is about 20 or larger
\cite{GailitisFreiberg:TheoryScrewDynamo,%
        GailitisFreiberg:Nonhomogeneous,%
        LupyanShukurov:Screw,%
        Leorat:Modelling%
}.
The field concentrates in a cylindrical shell of width
$\delta \RHO ={\cal O}(\Rm^{-1/4})$ (for $\Rm\gg1$)
for a continuous velocity field
\cite{RuzmaikinEtal:HydroScrew} and
$\delta \RHO ={\cal O}(\Rm^{-1/3})$ for a discontinuous velocity profile
\cite{Ponomarenko:Theory}, provided $m,k={\cal O}(1)$.
At distances from $\RHO_0$ larger than $\delta \RHO$, advective distortion of the
nonaxisymmetric magnetic field cannot be balanced by local
dynamo action. Therefore the magnetic field must be weaker than
in the resonance shell around $\RHO_0$ and decays
exponentially in $(\RHO{-}\RHO_0)^2$.
\citet{Gilbert:FastPonomarenko} has obtained
asymptotic solutions for the fastest mode, $m,k={\cal O}(\Rm^{1/3})$ for
continuous, and $m,k={\cal O}(\Rm^{1/2})$ for discontinuous velocity
fields.

The nonlinear behavior of the screw dynamo has been studied only recently in a paper
by \citet{BassomGilbert:Nonlinear}, who have carried out an asymptotic
analysis of the
nonlinear case in the limit $\Reynolds \gg \Rm \gg 1$, where
$\Reynolds$ is the kinematic Reynolds number.
This implies a small magnetic Prandtl number, $\Pm\equiv\Rm/\Reynolds \ll 1$.
The basic idea of their approach is that
the overall effect of the magnetic field on the flow is dominated by the
azimuthally and axially averaged Lorentz force.
The exponential growth of the kinematic stage is saturated via
a reduction in the velocity shear in the vicinity of $\RHO=\RHO_0$ where
dynamo action is most efficient.
In the asymptotic limit considered, the velocity shear is
fully suppressed (to a given asymptotic order)
by magnetic forces in a shell of a radial width
$\sim{\cal O}(\Rm^{-1/10})$.
Outside the shell, where the magnetic field is weaker,
magnetic diffusion and stretching balance each other
to maintain the magnetic field against Ohmic decay.
At still larger distances from $\RHO_0$, shear dominates
and the magnetic field is weak as at the kinematic stage.
The steady-state field strength $B_{\rm ss}$
in the spiral Couette flow is estimated as \cite{BassomGilbert:Nonlinear}
\begin{equation} \label{BG-Emag-Re}
  \frac{B_{\rm ss}^2}{\mu_0 \varrho U^2}
  = {\cal O}(\Rm^{2/5}\Reynolds^{-1}),
\end{equation}
where $U$ is a characteristic value of the velocity and $\mu_0$ denotes
the magnetic vacuum permeability.
As we argue below, the scaling with $\Reynolds$
is sensitive to the nature of the driving force and arises in
Eq.~(\ref{BG-Emag-Re}) because the flow is driven by viscous forces.

\citet{NunezEtal:Saturation} have performed classical perturbation
analysis close to the dynamo threshold for a system which is based on
Ponomarenko's discontinuous dynamo model \citep{Ponomarenko:Theory}, but
assuming the interior of the cylinder to contain fluid of fixed
viscosity.
For this sufficiently different system, they also find the asymptotic
scaling $B_{\rm ss}^2 \sim \Reynolds^{-1}$, while no conclusions on the
asymptotic dependence on $\Rm$ can be drawn from such a model.

\section{The model}
\label{Sec-Model}

\subsection{Spiral Couette flow}

\begin{figure}[t!]
  \centering
  \includegraphics[height=0.4\textwidth]{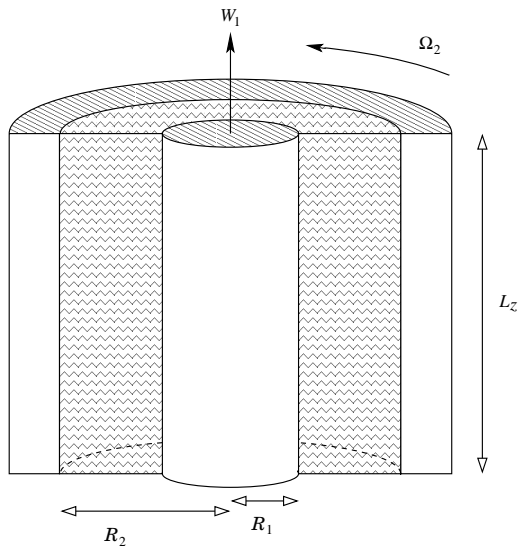}
  \caption{Geometrical configuration of the simulations. The outer cylinder
    rotates at angular velocity $\Omega_2$, while the inner cylinder moves
    in the axial direction at speed $W_1$.
  }
  \label{Fig-Geometry}
\end{figure}

The geometry of our model is shown in Fig.~\ref{Fig-Geometry}.
The conducting fluid is confined in the gap $R_1 < \RHO < R_2$
between two coaxial cylinders
that move with axial velocities $W_1$ and $W_2$ and rotate with
angular velocities $\Omega_1$ and $\Omega_2$, respectively.
We choose $\Omega_1=W_2=0$; the resulting flow
then trivially satisfies Rayleigh's stability criterion,
$\Omega_2 R_2^2>\Omega_1 R_1^2$
\citep{C1981}.
The magneto-rotational instability of the Couette flow is discussed in
Refs.~\cite{JGK2001,GJ2001,RZ2001}; our flow is stable with respect to
this instability because $d\Omega/d\RHO>0$.

In the absence of a magnetic field, viscosity causes the fluid between the
cylinders to adjust itself to the spiral Couette velocity profile
\begin{equation} \label{spiral-Couette}
  \Omega^{\rm(C)} = C_1 \biggl( 1 - \frac{R_1^2}{\RHO^2} \biggr),\qquad
  u_z^{\rm(C)}    = C_2 \ln \frac{R_2}{\RHO},
\end{equation}
where
\[
  C_1 = \frac{\Omega_2 R_2^2}{R_2^2{-}R_1^2},\qquad
  C_2 = \frac{W_1}{\ln (R_2/R_1)}.
\]

The velocity profile (\ref{spiral-Couette}),
driven by the viscous stress, adjusts itself over
the viscous relaxation time,
\begin{equation}  \label{tauv}
  \tau_{\rm visc} \approx \frac{(R_2{-}R_1)^2}{\pi^2\nu} .
\end{equation}

\begingroup
\begin{table*}[t!]
  \newcommand{\mcc}[1]{\multicolumn{1}{c}{#1}}
  \caption{Parameters of the numerical simulations discussed in the text.
    For all runs, $R_1 = 0.3$, $R_2=1.2$, $L_z=4$, $\Omega_1=0$, and $\Uz_2=0$.
    The values of $\Uz_1$, $\Omega_2$, $\nu $ and $\eta$ different from
    those of Model~1a are highlighted with bold face.
    The mesh has equal spacings in all three dimensions,
    $\delta x=\delta y=\delta z$ in all models except Model~1i
    where $\delta y=\delta x=\delta z/2$.
    Lengths are measured in units of $R_0$, angular velocities in
    $\Omega_0$, viscosity and diffusivity in $R_0^2\Omega_0$ and
    velocities in $R_0\Omega_0$.
  }
  \label{Tab-params}
  \begin{ruledtabular}
  \begin{tabular}{l@{\ecs}rrrrrrrrll}
    \multicolumn{1}{c}{Model}
       & \mcc{$\Uz_1$}
               & \mcc{$\Omega_2$}
                       & \mcc{$\nu$}
                                    & \mcc{$\eta$}
                                                 & \mcc{$U$}
                                                          & \mcc{$U_{\rm max}$}
                                                                   & \mcc{$\Reynolds$}
                                                                            & \mcc{$\Rm$}
                                                                                    & \mcc{$\Pm$}
                                                                                                & \mcc{$\delta x$} \\
    \colrule
    1a & $1.0$ & $1.3$ & $2\EE{-2}$ & $7\EE{-3}$ & $1.85$ & $1.56$ & $111$  & $318$ & $\dig2.9$ & $0.033$ \\
    1b & $1.0$ & $1.3$ & {\mbf $7\EE{-2}$}
                                    & $7\EE{-3}$ & $1.85$ & $1.55$ & $31.8$ & $318$ & $10$      & $0.067$ \\
    1c & $1.0$ & $1.3$ & {\mbf $1\EE{-2}$}
                                    & $7\EE{-3}$ & $1.85$ & $1.55$ & $222$  & $318$ & $\dig1.4$ & $0.067$ \\
    1d & $1.0$ & $1.3$ & {\mbf $2.5\EE{-3}$}
                                    & $7\EE{-3}$ & $1.85$ & $1.55$ & $889$  & $318$ & $\dig3.6$ & $0.033$ \\
    1e & $1.0$ & $1.3$ & {\mbf $1\EE{-3}$}
                                    & $7\EE{-3}$ & $1.85$ & $1.55$ & $2220$ & $318$ & $\dig0.14$& $0.017$ \\
    1f & $1.0$ & $1.3$ & $2\EE{-2}$ & {\mbf $8\EE{-3}$}
                                                 & $1.85$ & $1.55$ & $111$  & $278$ & $\dig2.5$ & $0.067$ \\
    1g & $1.0$ & $1.3$ & $2\EE{-2}$ & {\mbf $4\EE{-3}$}
                                                 & $1.85$ & $1.55$ & $111$  & $555$ & $\dig5.0$ & $0.067$ \\
    1h & $1.0$ & $1.3$ & $2\EE{-2}$ & {\mbf $2\EE{-3}$}
                                                 & $1.85$ & $1.56$ & $111$  & $1110$& $10$      & $0.033$ \\
    1i & $1.0$ & $1.3$ & $2\EE{-2}$ & {\mbf $1\EE{-3}$}
                                                 & $1.85$ & $1.56$ & $111$  & $2220$& $20$      & $0.017$ \\
    \colrule
    2a & {\mbf $0.5$}
               & $1.3$ & $2\EE{-2}$ & {\mbf $4\EE{-3}$}
                                                 & $1.64$ & $1.55$ & $98$   & $491$ & $\dig5.0$ & $0.033$ \\
    2b & {\mbf $0.5$}
               & $1.3$ & {\mbf $1\EE{-2}$}
                                    & {\mbf $4\EE{-3}$}
                                                 & $1.64$ & $1.55$ & $197$  & $491$ & $\dig2.5$ & $0.067$ \\
    2c & {\mbf $0.5$}
               & $1.3$ & {\mbf $7\EE{-2}$}
                                    & {\mbf $4\EE{-3}$}
                                                 & $1.64$ & $1.55$ & $197$  & $491$ & $18$      & $0.067$ \\
    2d & {\mbf $0.5$}
               & $1.3$ & $2\EE{-2}$ & {\mbf $2\EE{-3}$}
                                                 & $1.64$ & $1.55$ & $98$   & $984$ & $10$      & $0.033$ \\
  \end{tabular}
\end{ruledtabular}
\end{table*}
\endgroup


\subsection{Basic equations}

The equations we solve are the
MHD equations for the vector potential $\Av$, the
velocity $\uv$ and the density $\varrho$:
\begin{eqnarray}
  \label{PDE}
  \label{PDE-Av}
  \frac{\partial \Av}{\partial t}
  &=&
  \uv\,{\times}\Bv + \eta\Delta\Av + (\Div\Av)\grad\eta,\\[1ex]
  \label{PDE-uv}
  \frac{D\uv}{Dt}
  &=&
  -\frac{1}{\varrho}\grad p
  + \frac{\mu}{\varrho}\left(\Delta\uv+{\textstyle\frac13}\grad\Div\uv\right)
  + \frac{\jv{\times}\Bv}{\varrho},
\\[1ex]\
  \label{PDE-rho}
  \frac{D\varrho}{Dt}
  &=&
  - \varrho\Div\uv,
\end{eqnarray}
complemented by an isothermal equation of state,
$p=\csound^2\varrho$, with
constant speed of sound $\csound$.
Here, the magnetic flux density $\Bv$ and electric current density $\jv$
are given by
$\Bv = \rot\Av$, $\mu_0\jv = \rot\Bv$;
we denote with
$D/Dt \equiv \partial/\partial t + (\uv\cdot\grad)$
the advective derivative;
$\mu$ is the dynamical viscosity (assumed constant).
Below we refer to the Reynolds number based
on the average
kinematic viscosity, $\nu=\mu/\overline{\varrho}$.

Equation (\ref{PDE-Av}) implies the gauge
\begin{equation} \label{Gauge}
  \eta\Div\Av + \Phi = 0,
\end{equation}
where $\Phi$ is the electrostatic potential, related
to the electric field $\Ev$ by
$\Ev =-\grad\Phi-\partial\Av/\partial t$.
This gauge proved to be most convenient for numerical purposes.
Equations (\ref{PDE-Av})--(\ref{PDE-rho}) are written for compressible
fluids, but our choice of parameters makes compressibility insignificant
since the speed of sound is a factor of two larger than the maximum fluid
velocity, which results in a density contrast of $\le12\%$.

We use an explicit finite-difference scheme of sixth
order in space and third order in time
described, e.g., in Ref.~\citep{SanchezBrandenburg:2001}.
The velocity field outside the
fluid shell, i.e.\ for
$\RHO<R_1$ and $\RHO > R_2$, is prescribed and fixed, with
$\uv=(0,0,W_1)$ in $\RHO<R_1$ and $\uv=(0,\Omega_2\RHO,0)$ in $\RHO>R_2$.
We embed the cylinders into a Cartesian
box and solve Eqs.~(\ref{PDE-Av})--(\ref{PDE-rho}) on a Cartesian mesh
in order to avoid a coordinate singularity on the axis
and to retain the applicability of the code to different geometries.

The magnetic diffusivity $\eta$ is assumed to be constant for
$\RHO < R_2-3\delta x$ (with $\delta x$ the mesh size),
i.e.~the inner cylinder has the same electric conductivity as the
fluid, but $\eta$
smoothly decreases to zero in
$R_2-3\delta x<\RHO<R_2$.
Thus, the last term in Eq.~(\ref{PDE-Av}) is only relevant
close to the outer boundary of the fluid.
The outer cylinder is assumed to be magnetically impenetrable,
which confines the magnetic field to the region $\RHO < R_2$.
This would best be achieved with a perfect conductor at $\RHO\geq R_2$, but
this corresponds to an infinite magnetic Reynolds number, the numerical
implementation of which leads to fundamental difficulties.
Therefore we use
the stronger requirement $\Av=\zerov \text{ for } \RHO \ge R_2$ instead.
We demonstrate in Sect.~\ref{Sec-Linear-Results} that our
results are consistent with those obtained with a perfectly conducting
outer cylinder
since the magnetic field tends to concentrate close to the inner radius
$R_1$ and thus the outer boundary condition only weakly affects
the solution.
We have counter-checked our results with a modified magnetic
condition, where the vector potential is `softly' set to zero in the
region $\RHO>R_2$ by means of an additional relaxation term $-\Av/\tau_{\Av}$ in the
induction equation~(\ref{PDE-Av}), and we only report results that are not
qualitatively affected by this change.

In all three directions, periodic boundary conditions are assumed, which
are imposed on the faces of the computational box.
The horizontal boundary conditions are not actually important, since both the
fluid and the magnetic field are confined to $\RHO < R_2$.
In the axial direction (the $z$-direction), the assumption of periodicity
introduces a maximum wavelength $L_z$ (the vertical size of the box) and
leads to a quantization of the vertical wavenumber $k$ to
\begin{equation}\label{quant}
  k_n = 2\pi n/L_z,
\end{equation}
with integer $n$, for solutions that are
harmonic in $z$.
To assess the role of this quantization, we have carried out one numerical
experiment with a four times larger length $L_z=16$, which introduces
additional modes, which can have values of $k$ four times
closer to each other.
The result was qualitatively quite similar to our other models.
The only significant difference was that the system now had two modes with
different $k$, but very similar growth rates, which introduced a
relatively slow drift of energy from one mode to the other.
We will not discuss this simulation in the following, but rather focus on
the case of one fixed value of $L_z$ to ensure the comparability of the
different models.

\begin{figure}[t!]
  \centering
  \includegraphics[width=0.45\textwidth]{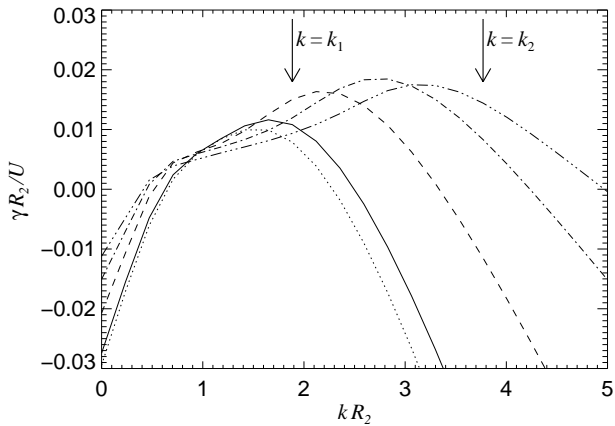}
  \caption{Kinematic growth rate $\gammar$ versus vertical
    wave number $k$ for Models
    1f ($\cdots$),
    1a (\protect\rule[0.6ex]{4ex}{0.2pt}),
    1g ($-\,-$),
    1h (\mbox{$- \cdot - \cdot{}$}).
    and 1i (\mbox{$- \cdot\cdot\cdot - \cdot\cdot\cdot{}$}).
    The first two wave numbers $k_1$, $k_2$ allowed by $L_z=4$ are
    indicated by vertical arrows.
  }
  \label{Fig-gamma-k}
\end{figure}

Figure~\ref{Fig-gamma-k} shows how the quantization of $k$ due to the
finite value of $L_z$ affects the dynamo system.
We show the dependence of the kinematic growth rate $\gammar$ on
the continuous wave number $k$ --- obtained by solving the eigenvalue
problem (\ref{Pono-1d-rho})--(\ref{Pono-1d-phi}) as described in
Sect.~\ref{Sec-Linear-Results} --- and indicate the quantized values
$k_n$ which occur in our simulations.
While the maximum growth rate can be up to 40\% larger than the maximum
rate measured at $k_n$ (Model 1d), the optimal values of $k$
do not differ much from $k_1$.
Thus, our choice of $L_z$ does not impose unrealistically small vertical
scales on the magnetic field.

\subsection{Parameters}
\label{Sec-Params}

The dynamo and hydrodynamical properties of the system are characterized
by two non-dimensional quantities, the kinematic and magnetic Reynolds
numbers, which are defined as
\begin{equation} \label{Re-Rm-def}
  \Reynolds = {U R_2}/{\nu}, \qquad
  \Rm = {U R_2}/{\eta},
\end{equation}
where $U=\sqrt{(R_2\Omega_2)^2 {+} \Uz_1^2}$ is a characteristic
velocity. The ratio of the two Reynolds numbers is the magnetic Prandtl number
\begin{equation}
  \Pm = \Rm/\Reynolds.
\end{equation}
We introduce a reference radius $R_0$ and a reference angular velocity
$\Omega_0$ such that $R_2=1.2R_0$ and $\Omega_2=1.3\Omega_0$, i.e.\ the
radial extent of the system and the revolution time are of order unity
when measured in units of $R_0$ and $\Omega_0$, respectively.
In the following, we will measure length in units of $R_0$, time in
units of $1/\Omega_0$ and velocity in units of $R_0\Omega_0$, without
explicitly indicating this in the text.

The parameter range investigated here is indicated in
Table~\ref{Tab-params}.
Parameters that remain constant for all the simulations are
the cylinder radii ($R_1=0.3$ and $R_2=1.2$), and the size of
the computational box: $L_z = 4$ for the vertical size and
$L_x = L_y = 2(R_2+3\delta x)$ for the horizontal dimensions.
The numerical resolution is either $42{\times}42{\times}60$ (corresponding
to $\delta x\approx0.067$) or $78{\times}78{\times}120$
grid points ($\delta x \approx0.033$)
except for Model~1i where the higher resolution of
$144\times144\times120$ ($\delta x \approx0.017$) was used.

For the parameters used, the viscous relaxation time (\ref{tauv})
is in the range 1--80 for the models discussed below.
For comparison, the growth time of the magnetic field
is greater than about 43 for the models presented, and the turnover
time is $2\pi/|\Omega_2{-}\Omega_1| \approx 4.8$.

Models~1 and 2 differ in the value of $W_1$ and
have therefore different velocity profiles. This results in different dynamo
efficiencies; for example, $\Rmstar=218$ in Model~1
and $\Rmstar=384$ in Model~2. Another important distinction is the
position of the magnetic field maximum, $\RHO_0$: in Model~1, the magnetic field
is localized closer to the inner cylinder than in Model~2
(see e.g.\ the insets in Figs.~\ref{Fig-gam-om-Rm-lin-1} and
\ref{Fig-gam-om-Rm-lin-2} or Fig.~\ref{Fig-B2-r-1d-3d} in Appendix~B).
The models are further subdivided (1a--f, 2a--d)
according to the values of $\nu$ and $\eta$ in order to explore
the effects of varying magnetic and kinematic Reynolds numbers.
Most of our models have $\Pm>1$, except Model 1d ($\Pm\approx0.36$) and
Model 1e ($\Pm = 0.14$).

\begin{figure}[t!]
  \centering
  \includegraphics[width=0.45\textwidth]{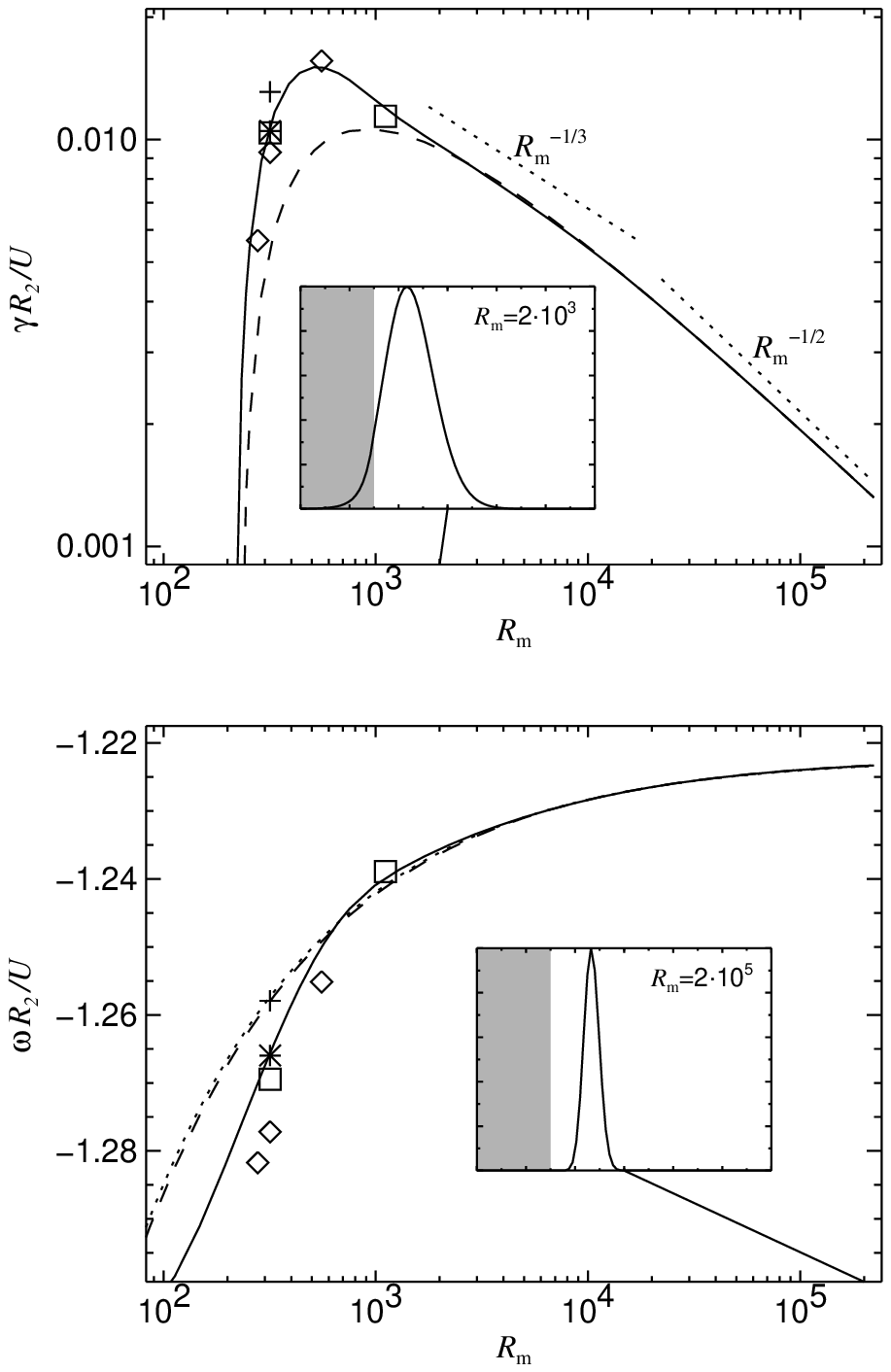}
  \caption{Kinematic growth rate $\gammar$ and frequency $\gammai$
    in Model~1
    for the mode $m=1$, $k=k_1$ (the dominant mode)
    as a function of the magnetic Reynolds number, obtained
    from the one-dimensional eigenvalue problem described in
    Sect.~\ref{Sec-Linear-Results}.
    Dotted lines show the
    dependencies $\gammar\propto\Rm^{-1/3}$ and
    $\gammar\propto\Rm^{-1/2}$ in the upper panel and
    $(\gammai-\mbox{const})\propto\Rm^{-1/2}$ in the
    lower panel.
    The asymptotic solution given
    in Appendix~\protect\ref{Sec-1d-asymp} is shown dashed.
    Results from the three-dimensional simulations are labeled according
    to the grid spacing:
    $\delta x = 0.13$ ($+$),
    $0.067$ ($\diamond$),
    $0.033$ ({\tiny$\Box$}),
    and $0.017$ ($*$).
    The insets show radial magnetic energy profiles in the domain
    $0<\RHO<R_2$ for $\Rm = 2\EE{3}$ (top) and
    $\Rm=2\EE{5}$ (bottom); the region occupied by the inner cylinder
    is shaded.
    The generation threshold for this configuration is $\Rmstar \approx 218$.
    }
  \label{Fig-gam-om-Rm-lin-1}
\end{figure}

\begin{figure}[t!]
  \centering
  \includegraphics[width=0.45\textwidth]{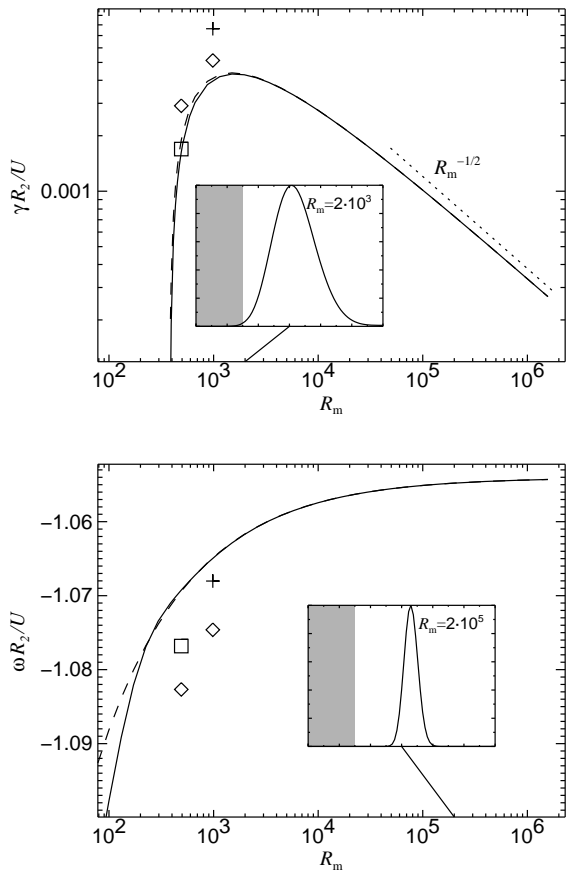}
  \caption{As in Fig.~\protect\ref{Fig-gam-om-Rm-lin-1}, but for Model~2.
    The critical magnetic Reynolds number is $\Rmstar \approx 384$.
  }
  \label{Fig-gam-om-Rm-lin-2}
\end{figure}

\section{Results}
\label{Sec-Results}

\subsection{Kinematic regime}
\label{Sec-Linear-Results}

The velocity profile (\ref{spiral-Couette}) can be considered as fixed, and
Eq.~(\ref{ind}) as linear in $\Bv$,
as long as the magnetic stress is weak compared to the viscous stress,
\begin{equation}\label{str}
  \frac{B^2}{\mu_0}
  \ll \varrho\nu\frac{U}{R_2{-}R_1}.
\end{equation}
Equations for the resulting kinematic dynamo problem are given in
Appendix~\ref{Sec-1d-kin-Eq}.
As discussed in Sect.~\ref{Sec-Linear-Theory},
they represent a one-dimensional eigenvalue problem which is relatively
straightforward to solve numerically.
For Models~1 and 2 we show in
Figs.~\ref{Fig-gam-om-Rm-lin-1} and \ref{Fig-gam-om-Rm-lin-2} the
dependence of the growth rate $\gammar$ and frequency $\gammai$ on
the magnetic Reynolds
number (solid lines) and compare them to the asymptotic formulae,
which are given in Ref.~\cite{RuzmaikinEtal:Couette-Poiseuille}
and are also reproduced in Appendix~\ref{Sec-1d-asymp} (dashed lines
in Figs.~\ref{Fig-gam-om-Rm-lin-1} and \ref{Fig-gam-om-Rm-lin-2}).
The insets show the radial profile of magnetic energy
for two values of $\Rm$.

The growth rate $\gammar$ becomes positive at $\Rm=\Rmstar$, first
quickly increases with $\Rm$, and then decreases as expected for a slow
dynamo \cite{ZRS1983,Soward:Unified}.
For the continuous velocity profile~(\ref{spiral-Couette}), the analytic
theory predicts an asymptotic decrease $\gammar\propto\Rm^{-1/2}$
(Appendix~\ref{Sec-1d-asymp}), and this scaling is indeed reached for very
large  values of $\Rm$.
However, while the growth rate for Model~2
(Fig.~\ref{Fig-gam-om-Rm-lin-2})
agrees well with the asymptotic result (dashed),
the agreement is not so good for $\Rm \lesssim 2000$ in Model~1
(Fig.~\ref{Fig-gam-om-Rm-lin-1}).
Moreover, the latter model shows a transient approximate scaling
$\gammar \sim \Rm^{-1/3}$.
Incidentally, this scaling is close to that for
a flow with discontinuous radial profile \cite{Ponomarenko:Theory,%
Gilbert:FastPonomarenko}.

The difference can be explained as follows.
For moderate magnetic Reynolds numbers, the
field has noticeable strength at the boundary of the inner
cylinder in the flow of Model~1 (see inset in the top frame of
Fig.~\ref{Fig-gam-om-Rm-lin-1}).
Therefore, the asymptotic theory is inapplicable as it is based
on the assumption that the magnetic field is concentrated far from the
boundaries $\RHO = R_1$ and $\RHO = R_2$.
However, the radial width of the field distribution decreases with $\Rm$
and eventually the field at $\RHO=R_1$ becomes negligible
(see inset
in the bottom frame of Fig.~\ref{Fig-gam-om-Rm-lin-1}),
and the scaling $\gammar \sim \Rm^{-1/2}$ is recovered.
On the other hand, the field is always small near the
boundaries in Model~2 (see insets in Fig.~\ref{Fig-gam-om-Rm-lin-2}),
and so the scaling $\gammar\sim\Rm^{-1/3}$ does not occur.

We have also explored the linear stage of magnetic field evolution
using the three-dimensional code in order to assess its performance.
A discussion of the numerical aspects is given in Appendix~\ref{app2}.
We initialize the simulations with a weak random magnetic field.
After the initial transients have died away,
exponential growth of magnetic energy is established,
corresponding to the fastest growing mode.

\begin{figure*}[t!]
  \centering
  \newlength{\explwidth}
  \setlength{\explwidth}{0.44\textwidth}
  \includegraphics[width=\explwidth]{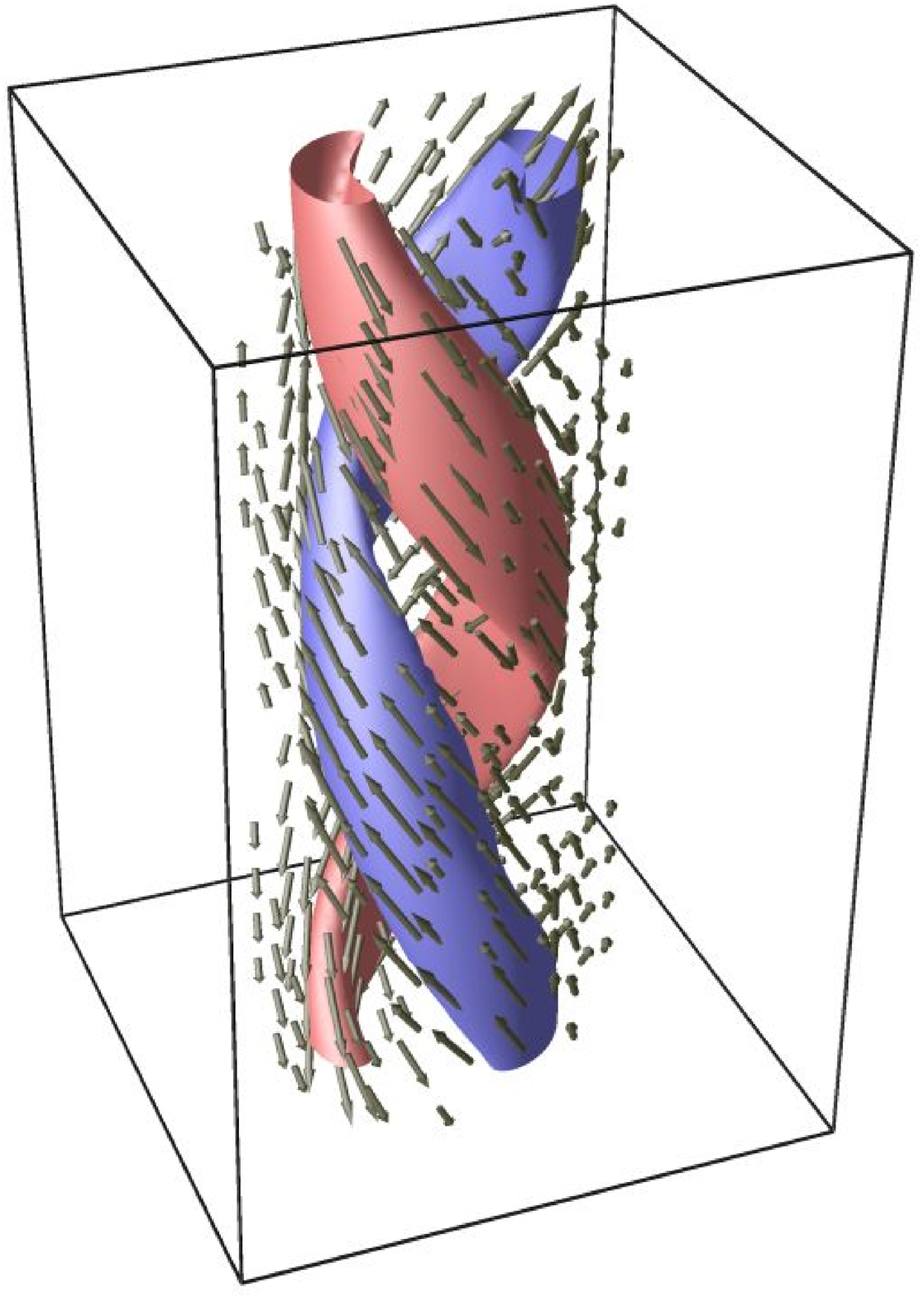}\hfill%
  \includegraphics[width=\explwidth]{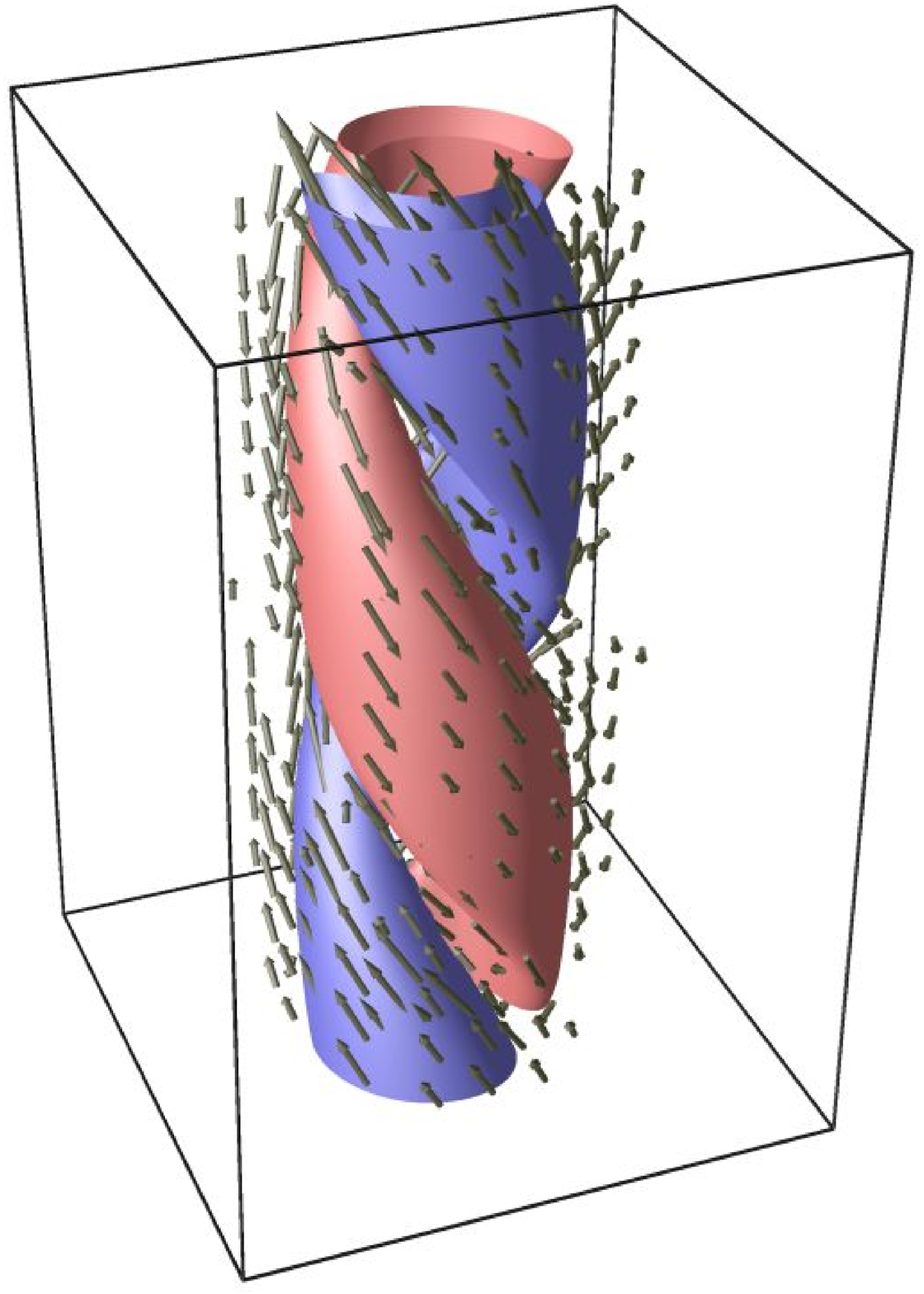}\\
  \makebox[0.44\textwidth][l]{\raisebox{1.36\explwidth}[0pt][0pt]{\ (a)}}\hfill%
  \makebox[0.44\textwidth][l]{\raisebox{1.36\explwidth}[0pt][0pt]{\ (b)}}\\
  \vspace*{-1\baselineskip}
  \caption{
    Three-dimensional representation of the magnetic field in Model~1h.
    (a) $t=200$ (end of the kinematic stage);
    (b) $t=912$ (the saturated state).
    Shown are surfaces where $|\Bv|$ is 0.65
    of the maximum and magnetic field vectors.
    The vertical component of the field has opposite sign in the
    two flux tubes shown (mode $m=1$).
    The azimuthal and axial variations of the field are harmonic at the
    kinematic stage (a) and flattened in the saturated state (b)
    (see also Fig.~\protect\ref{Fig-B-sat-3d}).
  }
  \label{Fig-Expl-1}
\end{figure*}

Figure~\ref{Fig-Expl-1}a shows the magnetic field structure for Model~1h.
The level surfaces of $|\Bv|$ have the form of two helical flux tubes
of opposite field orientation, which corresponds to an
azimuthal wavenumber $m=1$.
The vertical wavenumber of the solution shown is $k = 2\pi/L_z = \pi/2$.
This mode is excited in all the models of Table~\ref{Tab-params}.
However, additional higher modes are excited in Models~1h, 1i, and 2d,
where $\Rm$ is larger (see Table~\ref{Tab-results-nonlin}).

Both the flow and the magnetic field are strongly helical,
the two helicities being of opposite sign (the streamlines are
right-handed spirals, while the magnetic field lines form left-handed
helices as can be seen in Fig.~\ref{Fig-Expl-1}).
Since the screw dynamo mechanism relies on diffusion,
the approximate conservation of magnetic helicity in highly conducting media,
which leads to serious difficulties in mean-field dynamo theory
\cite{BF1999,Brandenburg:InverseCascade}, does not lead to any problem here.

\subsection{Nonlinear Simulations}
\label{Sec-Nonlin-Results}

After a phase of exponential growth, magnetic energy
levels off at a certain saturation value.
The corresponding magnetic energy density is comparable to (but smaller than)
the kinetic energy density in the sheared flow, as can be seen from
Table~\ref{Tab-results-nonlin}, where
we compare the maximum magnetic flux density to
$\sqrt{\overline{\varrho}U^2}$,
and the magnetic energy to
$E_{\rm kin} = \frac{1}{2}\overline{\varrho} U^2 V$, where
$V = L_z\pi(R_2^2{-}R_1^2)$ is the fluid volume.
The location and width of the magnetic energy maximum are characterized by
\begin{equation} \label{Def-r0eff}
    \RHO_{B}
    = \left< \RHO \right>_{\Bv^2} ,
    \qquad
    \delta\RHO_{B}
    = \left< (\RHO{-}\RHO_{B})^2 \right>_{\Bv^2}^{1/2} ,
\end{equation}
where $\left< f \right>_{\Bv^2} = \int f\, \Bv^2\,dV/\int \Bv^2\,dV$
is the magnetic-energy weighted volume average. These quantities are similar to $\RHO_0$
and $\delta \RHO$ of the kinematic theory.

\begingroup
\begin{table*}[t!]
  \centering
  \newcommand{\mcc}[1]{\multicolumn{1}{c}{#1}}
  \newcommand{\mcctwo}[1]{\multicolumn{2}{c}{#1}}
  \caption{
    Magnetic field properties in the models of
    Table~\ref{Tab-params}.
    Numbers in parentheses refer to the kinematic stage, all other
    quantities are for the saturated state.
    Shown are the growth rate $\gammar$ and
    oscillation frequency $\gammai$ of the leading mode;
    the maximum saturated magnetic field strength $B_{\rm max}$ and magnetic
    energy $E_{\rm mag}$ in the flow region, both normalized to the
    corresponding kinetic quantities; and the
    position and width of the magnetic field distribution in radius,
    $\RHO_{B}$ and $\delta\RHO_{B}$.
    The last column lists the growing modes in the form
    $(m,kL_z/(2\pi))$, $L_z=4$, ordered by decreasing growth rate.
    The results for Models~1h, 1i, and 2d refer to the $(1,1)$ mode only.
    Growth rates and oscillation frequencies are measured in units of
    $\Omega_0$, lengths in $R_0$.
  }
  \label{Tab-results-nonlin}
  \begin{ruledtabular}
  \begin{tabular}{c@{\quad}c@{\quad}c@{\,}c@{\quad}c@{\quad}l@{\quad}c@{\,}c@{\quad}c@{\,}c@{\quad}l}
    \multicolumn{1}{l}{Model}
        & {$\gamma$}
                     & \mcctwo{$\gammai$}
                                              & {$\dispfrac{B_{\rm max}}{\sqrt{\overline{\varrho}U^2}\strut}$}
                                                        & {$\dispfrac{E_{\rm mag}}{E_{\rm kin}}$}
                                                                   & \mcctwo{$\RHO_{B}$}
                                                                                         & \mcctwo{$\delta\RHO_{B}$}
                                                                                                              & $(m,2k/\pi)$ \\
  \colrule
    1a  & ($0.0160$) & $-2.03\dig$
                                 & ($-1.957$) & $0.33$  & $0.015$  & $0.49$  & ($0.47$)  & $0.17$  & ($0.16$) & $(1,1)$ \\
    1b  & ($0.0144$) & $-2.014$  & ($-1.957$) & $0.51$  & $0.034$  & $0.48$  & ($0.47$)  & $0.17$  & ($0.16$) & $(1,1)$ \\
    1c  & ($0.0144$) & $-2.037$  & ($-1.957$) & $0.24$  & $0.008$  & $0.48$  & ($0.47$)  & $0.17$  & ($0.16$) & $(1,1)$ \\
    1d  & ($0.0144$) & $-2.045$  & ($-1.957$) & $0.14$  & $0.002$  & $0.50$  & ($0.47$)  & $0.17$  & ($0.16$) & $(1,1)$ \\
    1e  & ($0.0144$) & $-2.051$  & ($-1.957$) & $0.08$  & $0.0009$ & $0.51$  & ($0.47$)  & $0.17$  & ($0.17$) & $(1,1)$ \\
    1f  & ($0.0087$) & $-2.014$  & ($-1.976$) & $0.25$  & $0.008$  & $0.48$  & ($0.47$)  & $0.17$  & ($0.17$) & $(1,1)$ \\
    1g  & ($0.0232$) & $-2.028$  & ($-1.935$) & $0.42$  & $0.024$  & $0.47$  & ($0.47$)  & $0.15$  & ($0.14$) & $(1,1)$ \\
    1h  & ($0.0176$) & $-2.012$  & ($-1.91$)  & $0.47$  & $0.030$  & $0.47$  & ($0.46$)  & $0.13$  & ($0.12$) & $(1,1)$, $(1,2)$ \\
    1i  & (---)      & $-1.981$  & (---)      & $0.55$  & $0.035$  & $0.46$  & (---)     & $0.11$  & (---)    & $(1,2)$, $(1,1)$, $(2,3)$, $(2,1)$, $(2,2)$ \\
    \colrule
    2a  & ($0.0023$) & $-1.459$  & ($-1.471$) & $0.17$  & $0.006$  & $0.63$  & ($0.66$)  & $0.18$  & ($0.18$) & $(1,1)$ \\
    2b  & ($0.0039$) & $-1.459$  & ($-1.473$) & $0.23$  & $0.10$   & $0.65$  & ($0.67$)  & $0.21$  & ($0.20$) & $(1,1)$ \\
    2c  & ($0.0039$) & $-1.506$  & ($-1.488$) & $0.48$  & $0.41$   & $0.63$  & ($0.66$)  & $0.20$  & ($0.20$) & $(1,1)$ \\
    2d  & ($0.0071$) & $-1.506$ & ($-1.467$) & $0.29$  & $0.01$   & $0.61$  & ($0.66$)  & $0.17$  & ($0.17$) & $(1,1)$, $(1,2)$ \\
  \end{tabular}
  \end{ruledtabular}
\end{table*}
\endgroup

\subsubsection{Spatial structure of the magnetic field}
\begin{figure}[t!]
  \centering
  \includegraphics[width=0.45\textwidth]{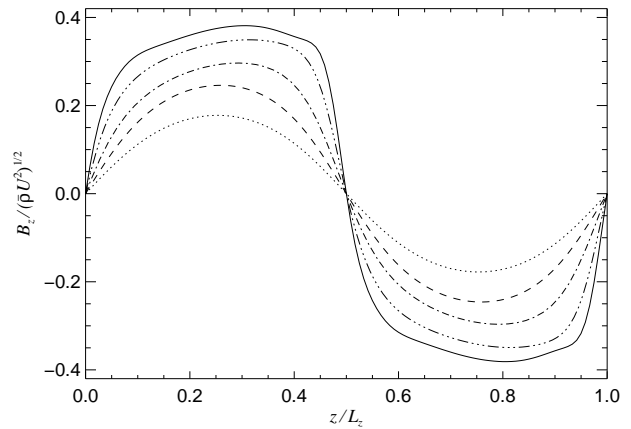}
  \caption{Magnetic field structure in the saturation regime.
    Shown is, as a function of $z$, the vertical field $B_z$ along a
    vertical line at $\RHO=0.48$ for
    Models 1f, 1a, 1g, 1h, and 1i;
    $\Rm$ grows monotonically along this sequence.
  }
  \label{Fig-B-sat-3d}
\end{figure}

Figure \ref{Fig-B-sat-3d} shows the
vertical profiles of the magnetic field in
the saturated regime for various magnetic Reynolds numbers.
The vertical component of the magnetic field
is plotted as a function of $z$ close to the radius
where the field concentrates.
Note that the antisymmetry of the curve with respect to the middle of the box
indicates that only Fourier components with odd vertical
wavenumbers are excited,
which can be understood from the structure of the nonlinear terms in
Eqs.~(\ref{PDE-Av})--(\ref{PDE-rho}).
The profile in the saturated state is flattened in comparison
to the (kinematic) eigenmode and
this effect becomes more pronounced as $\Rm$ increases.
The flattening of the maxima in both $z$ and $\varphi$
can also be seen in Fig.~\ref{Fig-Expl-1}, where the level surface
$|\Bv| = 0.65\,|\Bv|_{\rm max}$ is shown toward the end of the linear
phase (Fig.~\ref{Fig-Expl-1}a) and in the saturated state
(Fig.~\ref{Fig-Expl-1}b).

The strongly anharmonic profiles $B(z)$ found for large $\Rm$
do not occur in
\citeauthor{BassomGilbert:Nonlinear}'s \cite{BassomGilbert:Nonlinear}
theory, which predicts harmonic profiles in the saturated  state.
The difference may be due to the fact
that here $\Pm>1$, whereas Bassom \& Gilbert assume $\Pm\ll1$.
The sensitivity of the solution to the value of the magnetic Prandtl number
(if this is the true reason for the difference) is striking.

\begin{figure}[t!]
  \centering
  \includegraphics[width=0.45\textwidth]{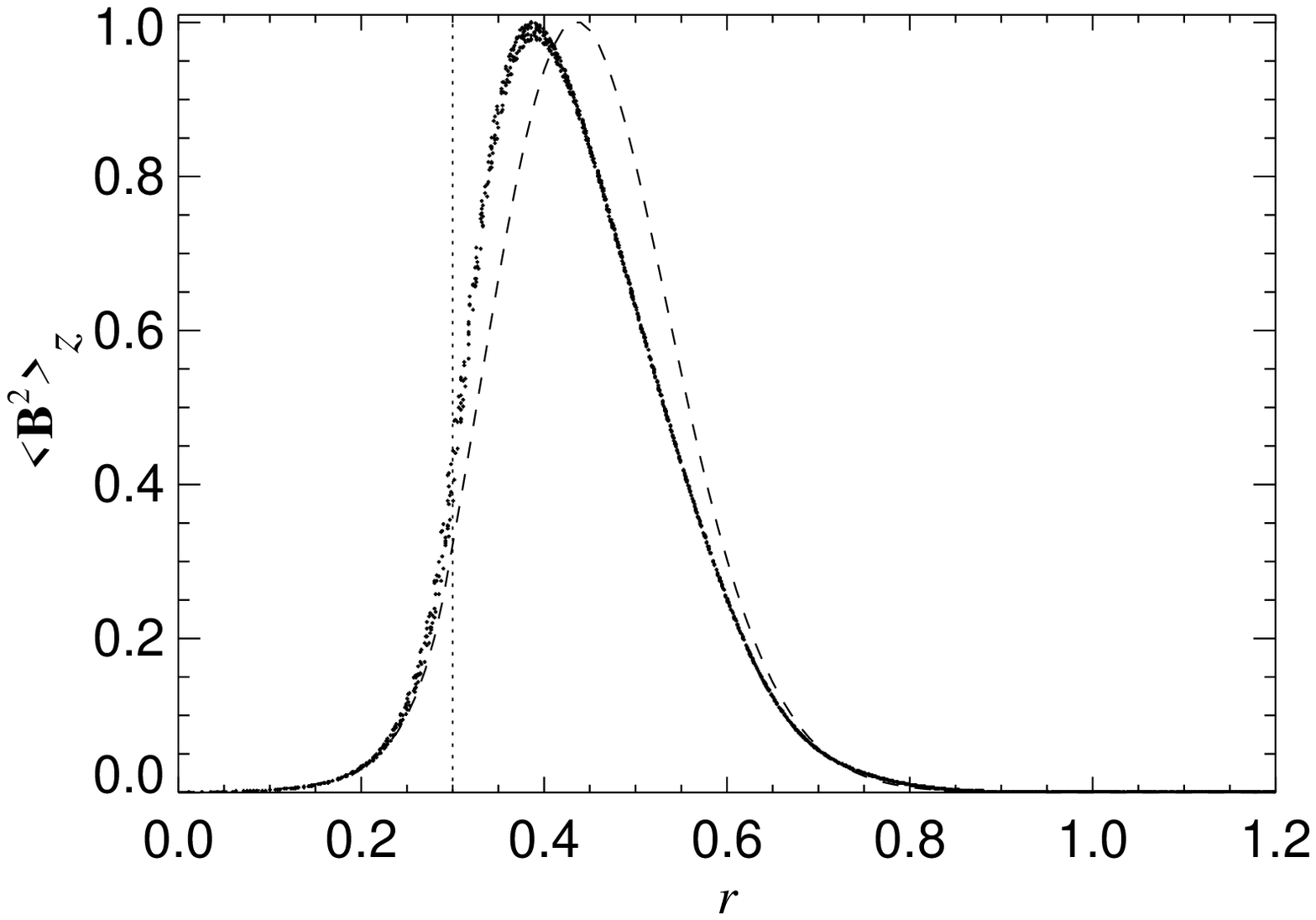}
  \includegraphics[width=0.45\textwidth]{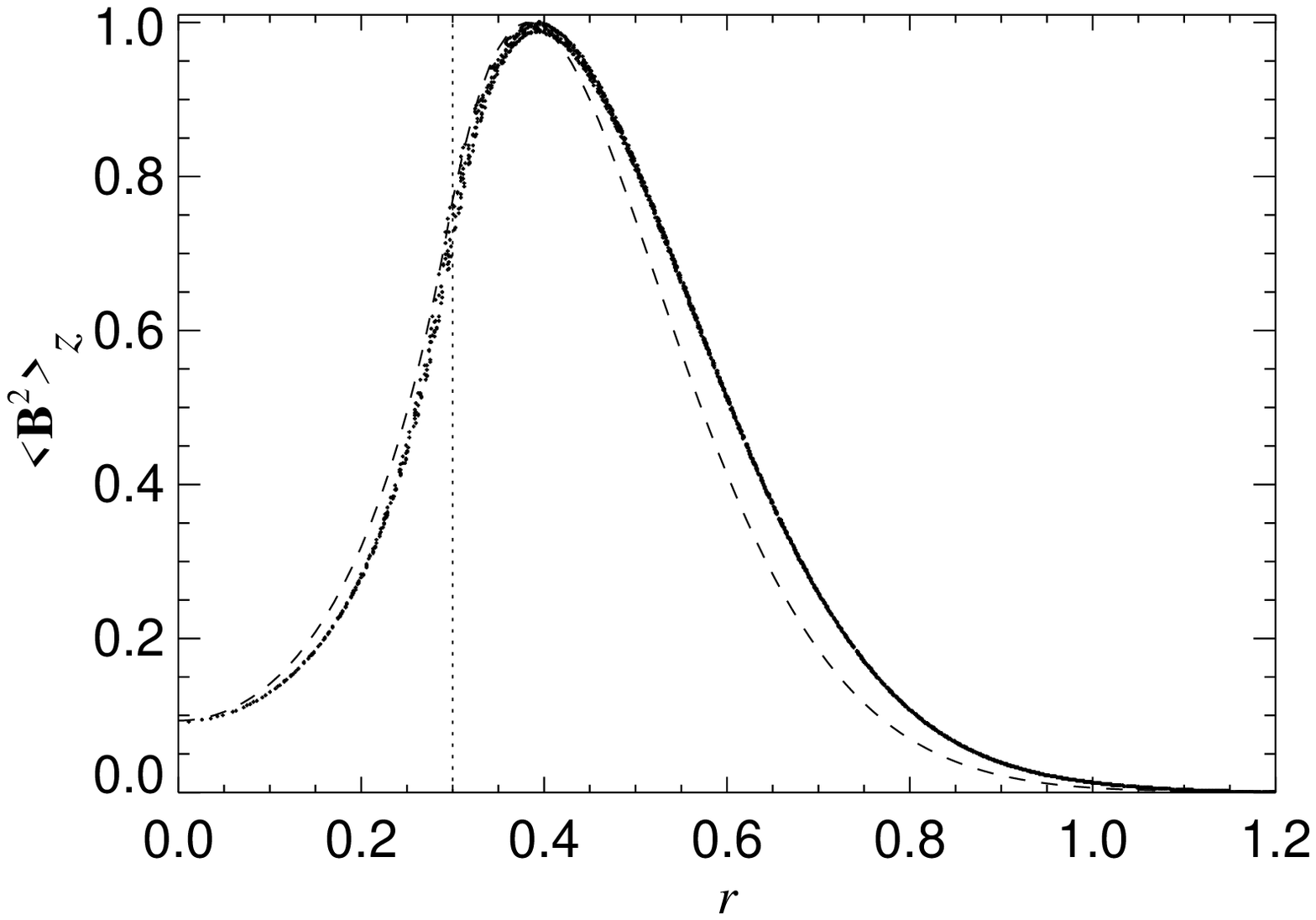}
  \caption{Radial magnetic field profile in the saturated regime for
    Models~1i (top; $\Pm=20$), and 1e (bottom; $\Pm=0.14$).
    Shown is the vertically averaged
    magnetic energy density as a function of $\RHO$ (dots) and the
    corresponding profile at the kinematic dynamo stage (dashed).
    The maximum of $\left<\Bv^2\right>_z$ is normalized to one; radius
    is measured in units of $R_0$.
  }
  \label{Fig-B2-r-sat}
\end{figure}

As can be seen in Fig.~\ref{Fig-B2-r-sat}, the nonlinear distortion of the
magnetic field distribution is only prominent in the
azimuthal and axial profiles, but not much in the radial profile.
We also note from Table~\ref{Tab-results-nonlin}
that in all the models
the radial width $\delta \RHO_B$ of the magnetic energy distribution in
the nonlinear stage is not significantly larger than that in the kinematic
stage.
In Model~1h, for example, the radial width of the eigenfunction is
$\delta\RHO_B \approx 0.12$ during the linear stage and $0.13$ at
saturation.
These numbers hardly differ and are very close to
$\delta\RHO$ from the asymptotic theory, which equals
$\delta \RHO=R_2 {\cal O}(\Rm^{-1/3})\simeq0.11$ for a discontinuous
velocity profile and $\delta \RHO=R_2 {\cal O}(\Rm^{-1/4})\simeq0.20$ for a
smooth profile.
In all our simulations, the radial width of the magnetic field
distribution in both linear and nonlinear states are very close to each
other and also close to that predicted by the asymptotic theory for the
kinematic dynamo.
This is at variance with the nonlinear asymptotics
of \citet{BassomGilbert:Nonlinear} that predict the development of a `core'
region in the radial profile of magnetic field with a width of order
${\cal O}(\Rm^{-1/10})\simeq0.5$ in Model~1i.
A possible reason for this discrepancy might be
that our models have too small values of $\Rm$ (2220 in Model~1i and 984 in
Model~2d) for the asymptotic regime of Bassom \& Gilbert to apply,
even though the linear asymptotics are already accurate.
It is more plausible, however, that
solutions with moderate magnetic Prandtl number do not develop the $\Rm^{-1/10}$
core and their radial profile is quite similar to that of the kinematic
eigenfunction (see Fig.~\ref{Fig-B2-r-sat}).

\begin{figure}[t!]
  \centering
  \includegraphics[width=0.4\textwidth]{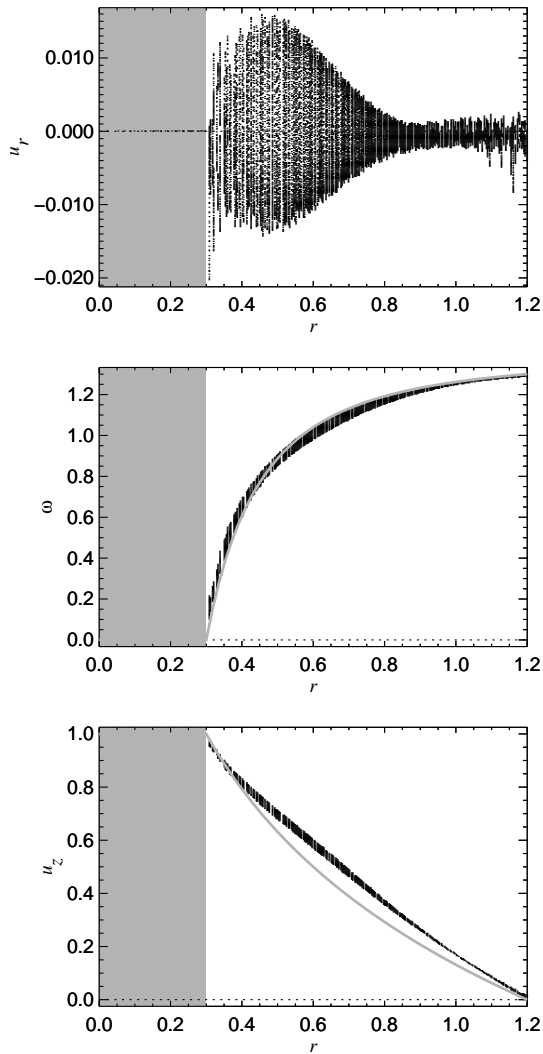}%
  \caption{Radial, azimuthal and vertical velocity
    components as functions of radius in the saturated stage for Model~1a
    at a resolution $\delta x = 0.033$.
    Note the different ranges on the ordinate in the panels:
    the variation of $u_\RHO$ is much smaller than that of
    $u_\varphi$ or $u_z$.
    The shear in both $\omega(\RHO)$ and $u_z(\RHO)$ is reduced
    in comparison to that in
    the Couette profile $\uv^{\rm(C)}$ of Eq.~(\ref{spiral-Couette}),
    which is shown as a continuous gray line.
    The location of the inner cylinder is marked in gray.
    The scatter of the data points arises from their different positions
    in $\varphi$ and $z$.
    Velocities are measured in units of $R_0\Omega_0$, angular velocities
    in units of $\Omega_0$ and radius in units of $R_0$.
    (In units of $U$ and $R_2$, the value $u_r=0.01$ corresponds to
    $0.0054U$, and $\omega=1.2$ corresponds to $0.78U/R_2$, for example.)
}
  \label{Fig-u-phi-z}
\end{figure}

\begin{figure}[t!]
  \centering
  \includegraphics[width=0.4\textwidth]{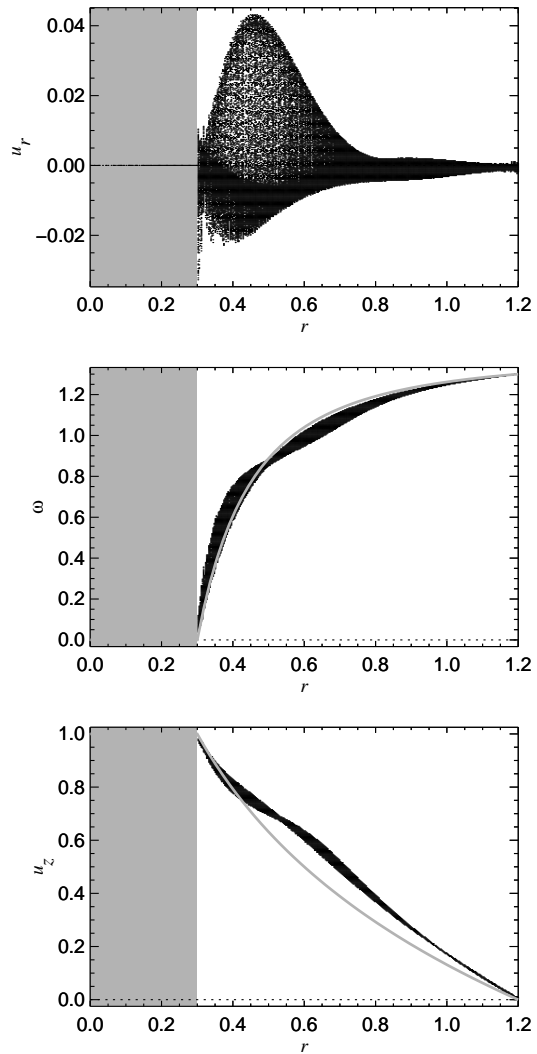}%
  \caption{As Fig.~\ref{Fig-u-phi-z}, but for Model~1i.
  }
  \label{Fig-u-phi-zd}
\end{figure}

\subsubsection{Nonlinear distortion of the velocity field}
\begin{figure}[t!]
  \centering
  \includegraphics[width=0.45\textwidth]{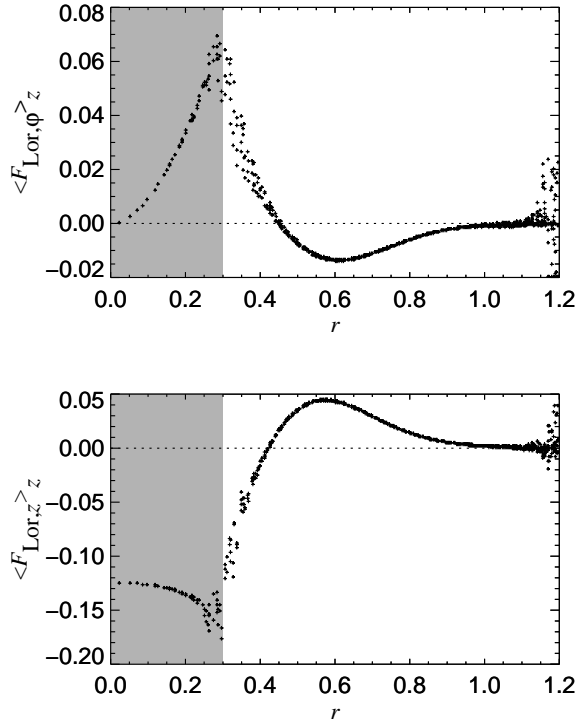}%
  \caption{Vertically averaged components of the Lorentz acceleration
  $\FLor=\jv\times\Bv/\varrho$ as a
  function of radius for Model~1a.
  Note that the effects of magnetic torques on the inner cylinder
  (shaded area in Fig.~\ref{Fig-F-Lor})
  are ignored in our model and its velocity remains fixed.
  Velocities are measured in units of $R_0\Omega_0$, angular velocities
  in units of $\Omega_0$ and radius in units of $R_0$.
  $\FLor$ is measured in units of $R_0\Omega_0^2$, radius in units of
  $R_0$.
  (In units of $U$ and $R_2$, the value $F_{\rm Lor}=0.05$ corresponds to
  $0.018U^2/R_2$, for example.)
  }
  \label{Fig-F-Lor}
\end{figure}

Figure~\ref{Fig-u-phi-z} shows the radial velocity profiles
for the saturated phase of Model~1a where the scatter of the data points is due
to their different positions in $\varphi$ and $z$.
The radial velocity fluctuates around zero and is dynamically unimportant
(even more so as perturbations in the radial velocity can be
balanced by the pressure gradient). However, the
azimuthal and axial velocities exhibit systematic deviations from their Couette
profiles so that
the velocity shear is reduced in the region where the magnetic field
concentrates.
It is especially clear in the case of $u_z$ (Fig.~\ref{Fig-u-phi-z}c) that
the spatial scatter is smaller than the mean variation, so the distortion
of the velocity field is axially symmetric and independent of $z$ to a first
approximation.
To justify this we demonstrate in
Fig.~\ref{Fig-F-Lor} that the radial profiles of the averaged Lorentz
force are in close correspondence with the
deviations of the respective velocity components of Fig.~\ref{Fig-u-phi-z}
from the Couette profile.
This is compatible with the scenario of \citet{BassomGilbert:Nonlinear},
where, in the limit of infinite kinematic and magnetic Reynolds numbers,
the saturation of dynamo action is mainly due to the
vertically and azimuthally averaged part
of the Lorentz force.
As the magnetic Reynolds number is increased, the relative distortion of
the velocity field, and in particular the reduction of velocity shear, are
getting more pronounced; this can be seen in Fig.~\ref{Fig-u-phi-zd},
where we show the same profiles as in Fig.~\ref{Fig-u-phi-z}, but for a
seven times larger value of $\Rm$.

The radial width of the region where the velocity field is distorted away
from the original Couette profile is much larger than that of the
Lorentz force (see Fig.~\ref{Fig-F-Lor} below).
This happens because the flow
adjusts itself to two separate Couette profiles at both ends of the
radial range where the Lorentz force has distorted it, and so a localized
magnetic field does affect the flow throughout the domain. It can be
expected that a flow profile driven by a volume force will be
distorted less in regions where magnetic field is weak; our results confirm
this expectation.

The screw dynamo can be interpreted in terms of the mean-field
$\alpha\Omega$-dynamo \cite{Soward:Unified,GilbertPonty:StreamSurfs},
the $\Omega$-term being as usual the shear term
$\RHO\Brho\,d\Omega/dr$ [see Eq.~(\ref{Pono-1d-phi})].
The $\alpha$-effect is identified with a part of the diffusion term, and
the corresponding term in the induction equation has the form
\begin{equation}
  - \eta \frac{2}{\RHO^2} \frac{\partial}{\partial\varphi} \Bphi
\end{equation}
[see also Eq.~(\ref{Pono-1d})].
Since the differential operator acting on $\Bphi$ is obviously not
influenced by the magnetic field strength,
the $\alpha$-effect cannot be affected by the growing magnetic field
and saturation is fully due to $\Omega$-quenching.
This is opposite to one of the standard scenarios for mean-field dynamos
($\alpha$-quenching), where the magnetic field has little influence on the
angular velocity and saturation
is considered to be caused by the partial
suppression of the $\alpha$-effect by magnetic fields.

\begin{figure}[t!]
  \centering
  \includegraphics[width=0.45\textwidth]{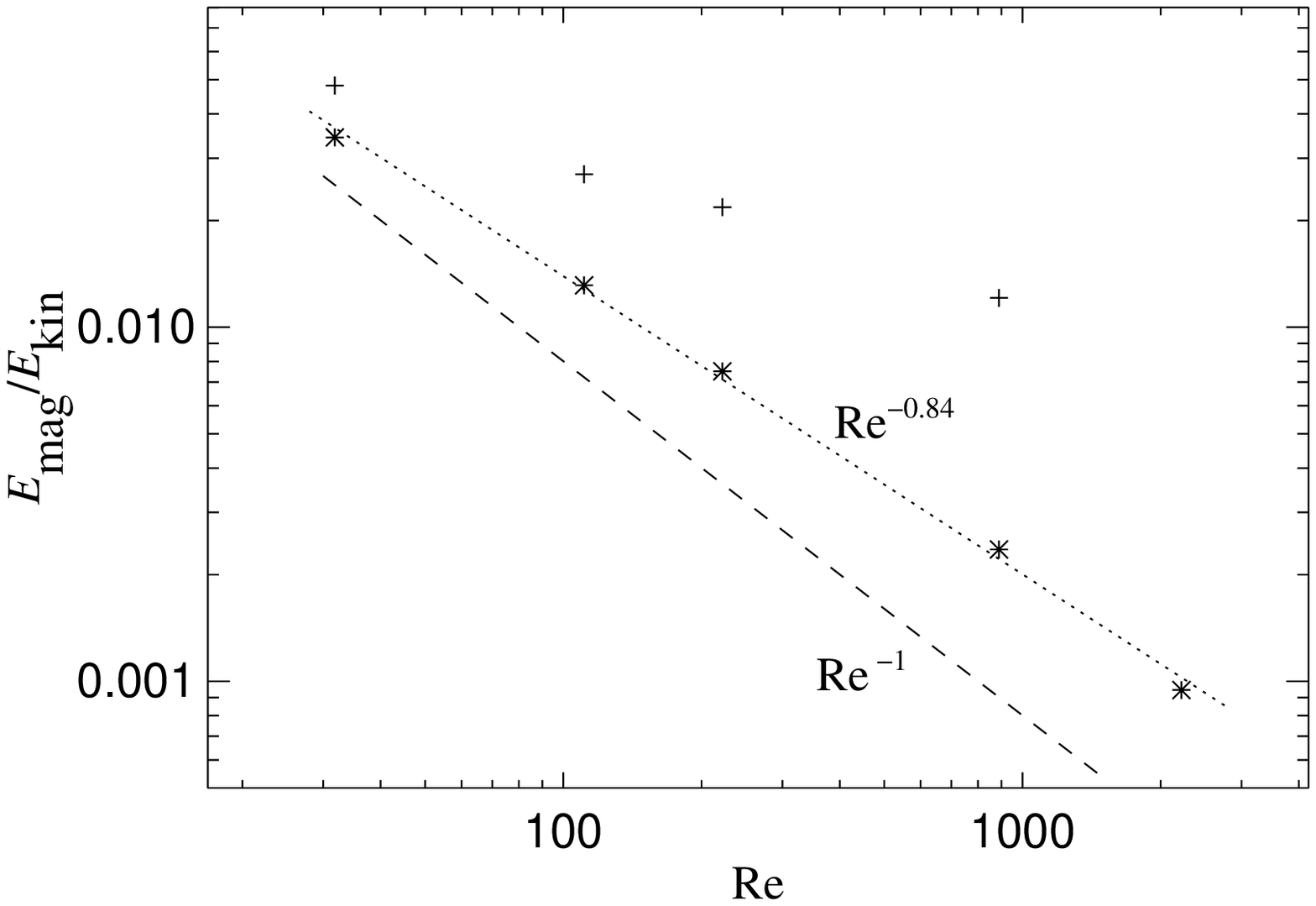}
  \includegraphics[width=0.45\textwidth]{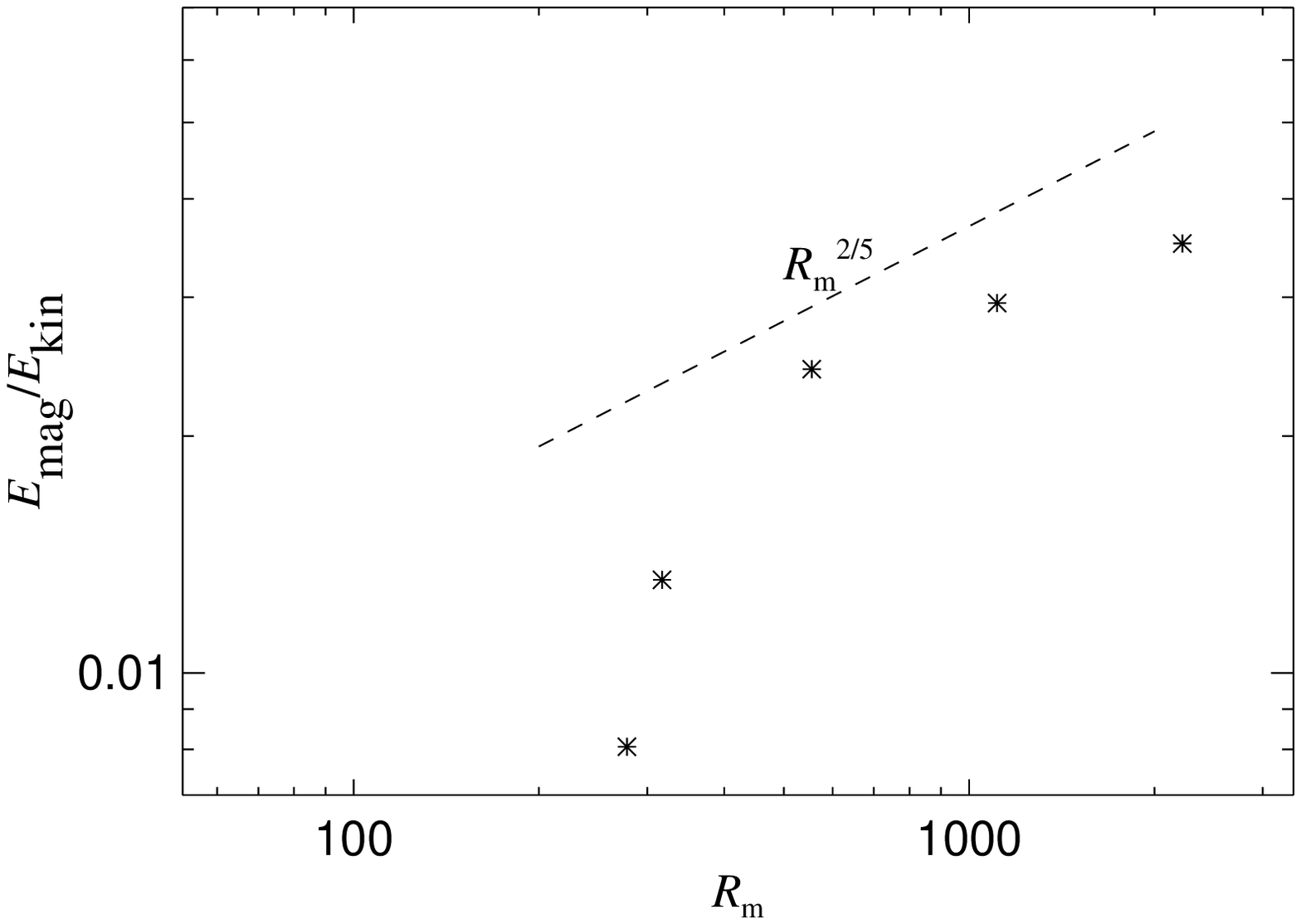}
  \caption{Magnetic energy in the saturated state versus the kinematic
    and magnetic Reynolds numbers for Models~1a--i.
    Top: $E_{\rm mag}/E_{\rm kin}$ as
    a function of $\Reynolds$ for Models~1b, 1a, 1c, 1d, and 1e (asterisks).
    The crosses refer to a model with volume forcing (see
    Sect.~\ref{Discussion-B-Re}).
    Bottom:  $E_{\rm mag}/E_{\rm kin}$ as a function of
    $\Rm$ for Models~1f, 1a, 1g, 1h, and 1i.
    The dashed lines show the asymptotic power laws given in
    Eq.~(\protect\ref{BG-Emag-Re});
    the dotted line is a least-square power-law fit and corresponds to a
    dependence $E_{\rm mag} \propto \Reynolds^{-0.84}$.
  }
  \label{Fig-Emag-Re}
\end{figure}

\subsubsection{Scaling of magnetic energy with $\Reynolds$ and $\Rm$}
Figure~\ref{Fig-Emag-Re} shows the magnetic energy
in the saturated state
as a function of the kinematic and magnetic Reynolds numbers. We also
show the asymptotic scalings of \citet{BassomGilbert:Nonlinear}.
There is clearly some agreement between the numerical and asymptotic results
although the asymptotic solution has been obtained for
$\Pm\ll1$ whereas the numerical results refer mostly to the
case $\Pm>1$.
We believe that the reason for the rough agreement is related
to the fact that, in both cases, the magnetic feedback affects mainly
the axisymmetric flow profile, as was seen in Fig.~\ref{Fig-u-phi-z}.

As expected (at least for weakly supercritical solutions) the
steady-state magnetic field strength $B_{\rm ss}$ increases with magnetic
Reynolds number.
On the other hand, the saturated field strength
$B_{\rm ss}$ decreases with $\Reynolds$ in a flow driven by viscous
stresses because it becomes easier for the magnetic field to modify the
velocity field in a given cylindrical shell as $\Reynolds$ increases
(owing to the weaker viscous coupling of
fluid at different radii), and so a weaker magnetic field is needed to achieve
a local reduction in shear sufficient to halt the field growth.
As we discuss below, this behavior is characteristic of a flow driven by
viscous stresses rather than by a volume force.

The scaling of magnetic energy with $\Reynolds$ is slightly shallower
than predicted by Bassom \& Gilbert and has an exponent close to $-0.84$ rather
than $-1$.
Similarly, the growth of magnetic energy with $\Rm$ is slower than
$\Rm^{2/5}$ at the values of $\Rm$ explored here.
The most plausible reason for these disagreements is the difference
in magnetic Prandtl numbers and, possibly, also the fact that the magnetic field
distribution is still not narrow compared to the gap width $R_2{-}R_1$ for
the magnetic Reynolds numbers we were able to consider.

\label{Discussion}

\label{Discussion-B-Re}
Taken at face value, the dependence of $B_{\rm ss}$ on $\Reynolds$ discussed above
implies that the resulting magnetic field
will be negligible wherever $\Reynolds\gg1$.
However,
in most real systems the flow will be driven by non-viscous forces. These
are pressure and inertial forces in the dynamo experiments in Riga and
Perm. In the case of
astrophysical jets, acceleration and collimation of the flow can be
due to an external magnetic field (axisymmetric to the first approximation, then
the screw dynamo can generate an additional nonaxisymmetric magnetic field).
In these cases the kinematic Reynolds
number will not play such a prominent role in the system and $B_{\rm ss}$
is expected to be independent of $\Reynolds$ for $\Reynolds \gg 1$.

To verify this idea, we have carried out further numerical
simulations for the parameters of Models~1b, 1a, 1c, and 1d, but with an
additional volume force
\begin{equation}
  -\frac{\uv{-}\uv^{\rm(C)}}{\tau}
\end{equation}
on the right-hand side of
the equation of motion (\ref{PDE-uv}).
Here $\uv^{\rm(C)}$ denotes the spiral Couette profile
(\ref{spiral-Couette}) and $\tau$ is the time scale over which the flow
adjusts itself to the Couette profile. With our choice $\tau=1$, we have
$\tau \lesssim \tau_{\rm visc}$, where the viscous time $\tau_{\rm visc}$ varies in
the range 1--80, cf.\ Eq.~(\ref{tauv}). Thus, the Couette flow profile
is now maintained on a dynamical time scale rather than by viscosity
if $\Reynolds\gg30$.

We show in Fig.~\ref{Fig-Emag-Re} by crosses the resulting dependence
of the steady-state magnetic energy on $\Reynolds$;
the dependence is clearly much weaker than in the case of
viscous driving (a power-law fit to the points shown has an exponent of
about $-0.41$).
In the inviscid limit $\Reynolds \to \infty$ the magnetic energy must
eventually become independent of viscosity,
and so we anticipate that the dependence marked with crosses has a horizontal
asymptote.

Another reason why the dependence (\ref{BG-Emag-Re}) is not directly
applicable to laboratory and astrophysical dynamos
is that the corresponding flows are turbulent.
In that case the generation of magnetic field on the scale
of the mean flow can be accompanied by the so-called `fluctuation
dynamo' producing small-scale magnetic fields, which eventually
achieves
energy equipartition with the turbulence.
For small magnetic Prandtl number, however, the dynamo is only weakly
supercritical and only large scales can be excited. Thus, the screw
dynamo effect is expected to work, but it is now controlled by the
effective values of $\Reynolds$ and $\Rm$ based on turbulent diffusivities.
We conjecture that this could still be qualitatively true in the
nonlinear phase of field evolution.
Since the turbulent values of $\Reynolds$ and $\Rm$ can be quite moderate,
even the scaling (\ref{BG-Emag-Re})
would not lead to physically uninteresting magnetic fields.
It can be expected, however, that the turbulence in the Couette flow ---
being generated mainly in turbulent boundary layers near the flow
boundaries --- will be very inhomogeneous and this can affect the theory
discussed here.
On the other hand, the
reduction in the effective magnetic Reynolds number due to turbulence has
to be only moderate since one needs $\Rm\geq\Rmstar$ for any sort of dynamo
action.
Assuming that turbulent magnetic and kinetic Reynolds numbers have
similar orders of magnitude, $\Reynolds\simeq\Rm$, and that $\Rmstar\simeq10^2$,
the resulting magnetic energy density will be about 10\% of the kinetic energy
density even with the scaling
(\ref{BG-Emag-Re}).

We note in this connection that the maximum radial magnetic field
component measured in the Riga dynamo experiment
\citep{GailitisEtal:Saturation} is about $6\,{\rm mT}$, while the
equipartition field strength would be $\approx0.5\,{\rm T}$.
$\Bphi$ and $\Bz$ being about five times larger than $\Brho$ in the
kinematic case [F.\ Stefani, private communication],
one gets a ratio $E_{\rm mag}/E_{\rm kin} \approx 5\EE{-3}$.
This is two orders of magnitude larger than $\Rm^{2/5}\Reynolds^{-1}
\lesssim 10^{-5}$ and clearly implies that the scaling (\ref{BG-Emag-Re}),
which is based on the assumption of a viscously driven, laminar flow,
cannot be directly applied to flows in laboratory experiments.

All our experiments were carried out for angular velocity $\Omega$
increasing outwards and, for the boundary conditions and parameters
considered, the flow was found to be hydrodynamically stable in the sense
that the Lorentz force led to relatively small deviations from the
unperturbed Couette profile (\ref{spiral-Couette}).
An interesting question is that of dynamo action in a spiral
Couette flow unstable with respect to the magneto-rotational instability,
i.e.\ for angular velocity decreasing outwards.
Flows of this type --- although for vanishing vertical velocities of the two
cylinders --- are currently discussed in connection with
experimental investigations of the magneto-rotational instability
\cite{JGK2001,GJ2001,RZ2001}.
If the flow is hydrodynamically stable (angular momentum increasing
outwards), a certain magnetic field is necessary to obtain regular Taylor
columns;
however, these columns involve a helical velocity field and are themselves
capable of dynamo action \cite{RZ2001}.
Thus, the magnetic field can play the role of a `catalyst', destabilizing
the system and at the same time regenerating it by the resulting
instability; this mechanism has been demonstrated to occur in accretion
discs \cite{BrandenburgEtal:DynamoGenerated,BalbusHawley:RevModPhys}.

\section{Conclusions}
\label{Sec-Conclusions}

The numerical calculations performed here have revealed a new intermediate
asymptotic
regime in the kinematic screw dynamo where, in a certain range of $\Rm$,
the magnetic field eigenfunction
is large enough at the flow boundary so that
the standard asymptotic solutions are inapplicable and the growth rate of
the magnetic field scales as $\gammar={\cal O}(\Rm^{-1/3})$.
This dependence is typical
of a discontinuous velocity profile. The standard asymptotic scaling,
$\gammar={\cal O}(\Rm^{-1/2})$, is only recovered at significantly larger
values of $\Rm$ where the eigenfunction becomes narrow enough in radius as
to satisfy the assumptions of the asymptotic theory.

We have confirmed the result of \citet{BassomGilbert:Nonlinear} that
saturation of screw dynamo action occurs via a reduction in the velocity
shear produced by the axisymmetric part of the Lorentz force.
We have shown that this also applies to the case of
large magnetic Prandtl numbers $\Pm$. However, the radial profile of the nonlinear
solution is very similar to the marginally stable eigenmode for the
nonlinearly modified velocity field. This is different from the nonlinear
asymptotics of Bassom and Gilbert, which predicts a plateau in the radial
dependence. It is possible that such a plateau can only occur at values
of the magnetic Prandtl number much smaller than what we have been able to
simulate in the present work,
but we have not detected any tendency towards its development at $\Pm=0.14$, the
smallest value explored here.

We have demonstrated that the scaling of the steady-state magnetic field with
kinematic Reynolds number is sensitive to the nature of the forces driving
the flow.

\appendix

\section{One-dimensional kinematic dynamo problem}

\label{Sec-1d-kin-Eq}

In cylindrical coordinates $(\RHO, \varphi, z)$,
the kinematic dynamo problem for given steady velocity field
$\bigl(0,\RHO\Omega(\RHO),u_z(\RHO)\bigr)$, is given by the following
non-dimensionalized set
of equations for the two amplitudes $\Bhatrho$, $\Bhatphi$
[cf.\ Eq.\ (\ref{B-harmon})]:
\begin{eqnarray} \label{Pono-1d}
  \label{Pono-1d-rho}
  \gammac \Bhatrho + {\rm i}(m\Omega{+}k u_z) \Bhatrho
  &=&
  \frac{1}{\Rm} \left\{
      \Dop\Bhatrho - \frac{2 {\rm i} m}{\RHO^2}\Bhatphi
                \right\} ,\\
  \label{Pono-1d-phi}
  \gammac \Bhatphi + {\rm i}(m\Omega{+}k u_z) \Bhatphi
  &=&
  \RHO \frac{d\Omega}{d\RHO} \Bhatrho \nonumber\\
        &&\mbox{}\!
  + \frac{1}{\Rm} \left\{
       \Dop\Bhatphi + \frac{2 {\rm i} m}{\RHO^2}\Bhatrho
                  \right\}\!\! ,
  \qquad
\end{eqnarray}
where
\begin{eqnarray}
  \Dop   &=& \frac{1}{\RHO} \frac{d}{d\RHO}
              \left( \RHO \frac{d}{d\RHO} \right)
              - \frac{m^2{+}1}{\RHO^2} - k^2
\end{eqnarray}
is a self-adjoint, Laplace-type differential operator,
and $\gammac$ is the eigenvalue.
The vertical component $B_z$ can be obtained from
 the solenoidality condition,
\begin{equation}
  \frac{1}{\RHO} \frac{d}{d\RHO} (\RHO\Bhatrho)
  + \frac{{\rm i} m}{\RHO} \Bhatphi
  + {\rm i}k\Bhatz
  = 0.
\end{equation}

\label{Sec-1d-asymp}

Second-order asymptotic analysis (for $\Rm \to \infty$) for the spiral
Couette flow (\ref{spiral-Couette}) has been presented in
Refs.\ \cite{RuzmaikinEtal:HydroScrew,RuzmaikinEtal:Couette-Poiseuille}.
It is convenient to define the magnetic Reynolds number as
\begin{equation}
  \Rmtilde = \frac{R_1^2 R_2^2}{R_2^2{-}R_1^2}
             \frac{|\Delta\Omega|}{\eta}
           \equiv 2\frac{|\Omega'(\RHO)|\RHO^3}{\eta}
\end{equation}
here, which is different from the definition (\ref{Re-Rm-def}) used in the
body of this paper.
The radius $\RHO_0$, where the magnetic field concentrates, is given by
\begin{equation}
  \RHO_0 = R_1 R_2 \sqrt{\frac{\ln R_2{-}\ln R_1}{R_2^2{-}R_1^2}}
                   \sqrt{-2\frac{m\Delta\Omega}{k\Delta W}}
\end{equation}
and the field can only be growing if
\begin{equation}
  \frac{m\Delta\Omega}{k\Delta W} < 0 ,
\end{equation}
where $\Delta\Omega=\Omega_2{-}\Omega_1$ and $\Delta W=W_2{-}W_1$.

To second order in the small parameter $\Rmtilde^{-1/2}$, the
eigenvalue $\gammac$
of the fastest growing mode is given (in dimensional units) by
\begin{eqnarray}
  \lefteqn{\gammac + {\rm i}[m\Omega(\RHO_0) + k u_z(\RHO_0)]} \nonumber\\
  &\quad=&
    \frac{\eta}{\RHO_0^2}
    \biggl\{
      (\sqrt{2}{-}1)\sqrt{|m|}
        (1{+}\sigma_1{\rm i})\Rmtilde^{1/2} \nonumber\\
  && \qquad{} + \Bigl(\frac{17}{36} -\sqrt{2}-k^2\RHO_0^2\Bigr)
     {} + {\cal O}\left( \Rmtilde^{-1/2} \right)
    \biggr\},\qquad
\end{eqnarray}
where $\sigma_1 = {\rm sgn}(k\Delta W)$ denotes the sign of $k\Delta W$.
As $\eta \propto \Rmtilde^{-1}$, the leading-order term on the right-hand
side is ${\cal O}\bigl( \Rmtilde^{-1/2} \bigr)$
when measured in the units $U/R_2$.

To the leading order in $\Rmtilde$, the magnetic field amplitudes
$\widehat{B}$ are given by
\begin{eqnarray}
  \Bhatphi &=& \exp\left[ -(1{+}\sigma_1{\rm i})\sqrt{|m|\Rmtilde}
                                 \frac{(\RHO{-}\RHO_0)^2}{2\RHO_0^2}
                              \right]
                          \nonumber\\&&{}
                + {\cal O}\left( \Rmtilde^{-1/2} \right) ,\qquad \\
  \Bhatrho &=& \sigma_2 (1{+}\sigma_1{\rm i})
                       \sqrt{\frac{|m|}{2}} \Bhatphi \Rmtilde^{-1/2}
             + {\cal O}\left( \Rmtilde^{-1} \right) ,    \\
  \Bhatz   &=& -\frac{m}{k}\left(\frac{1}{\RHO}
                                  + \sqrt{2}\,\frac{\RHO{-}\RHO_0}{\RHO_0^2}
                                \right) \Bhatphi
             + {\cal O}\!\left( \Rmtilde^{-1/2} \right)\! ,\qquad
\end{eqnarray}
where $\sigma_2={\rm sgn}\,\Delta\Omega$.

\section{Accuracy of the numerical scheme}  \label{app2}

We assess the accuracy of our three-dimensional simulations
and the implementation of the boundary conditions by
comparing solutions
obtained with the 3D code in the kinematic regime
with those from the corresponding
1D eigenvalue problem.

\begin{figure}[t!]
  \centering
  \includegraphics[width=0.45\textwidth]{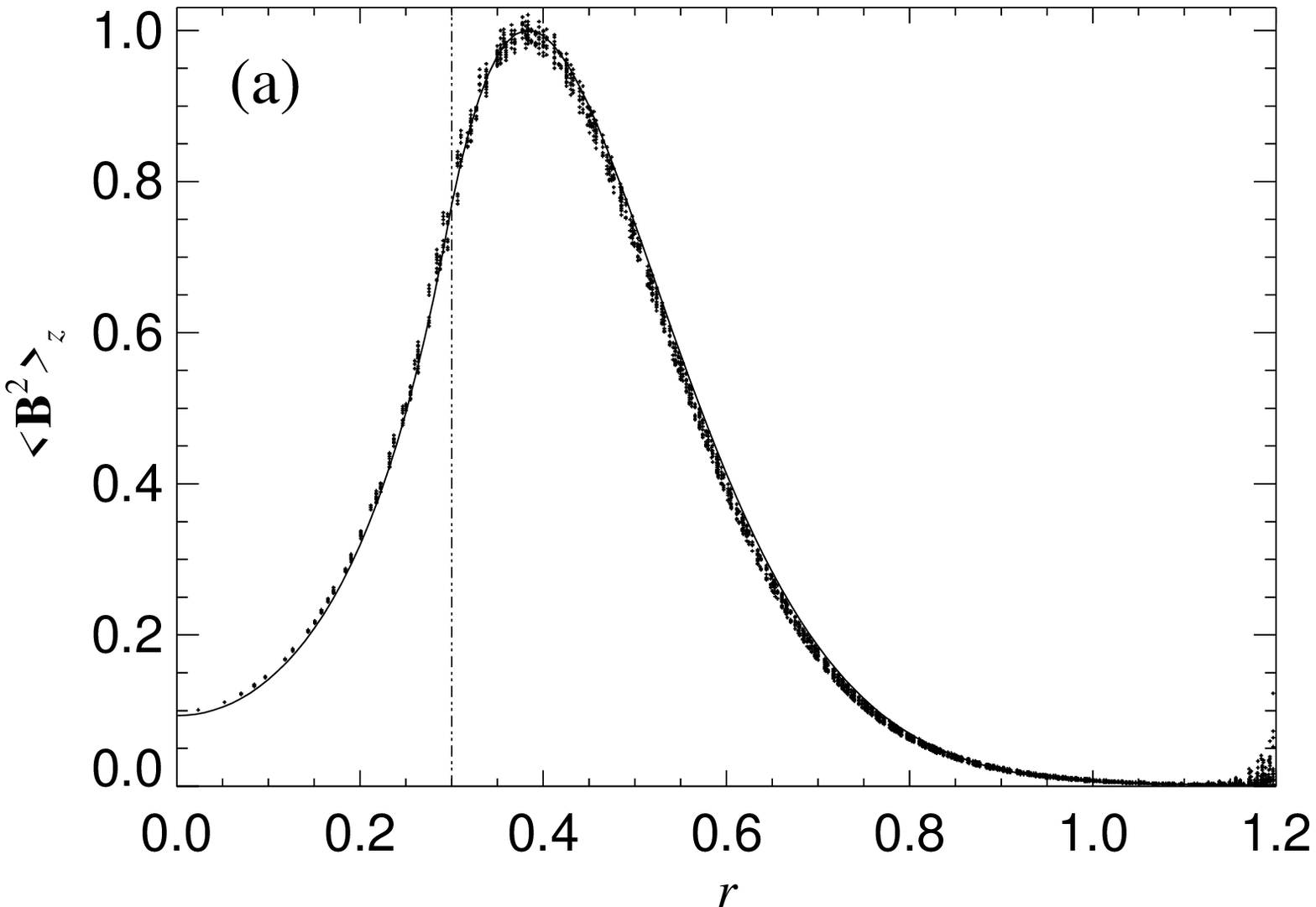}
  \includegraphics[width=0.45\textwidth]{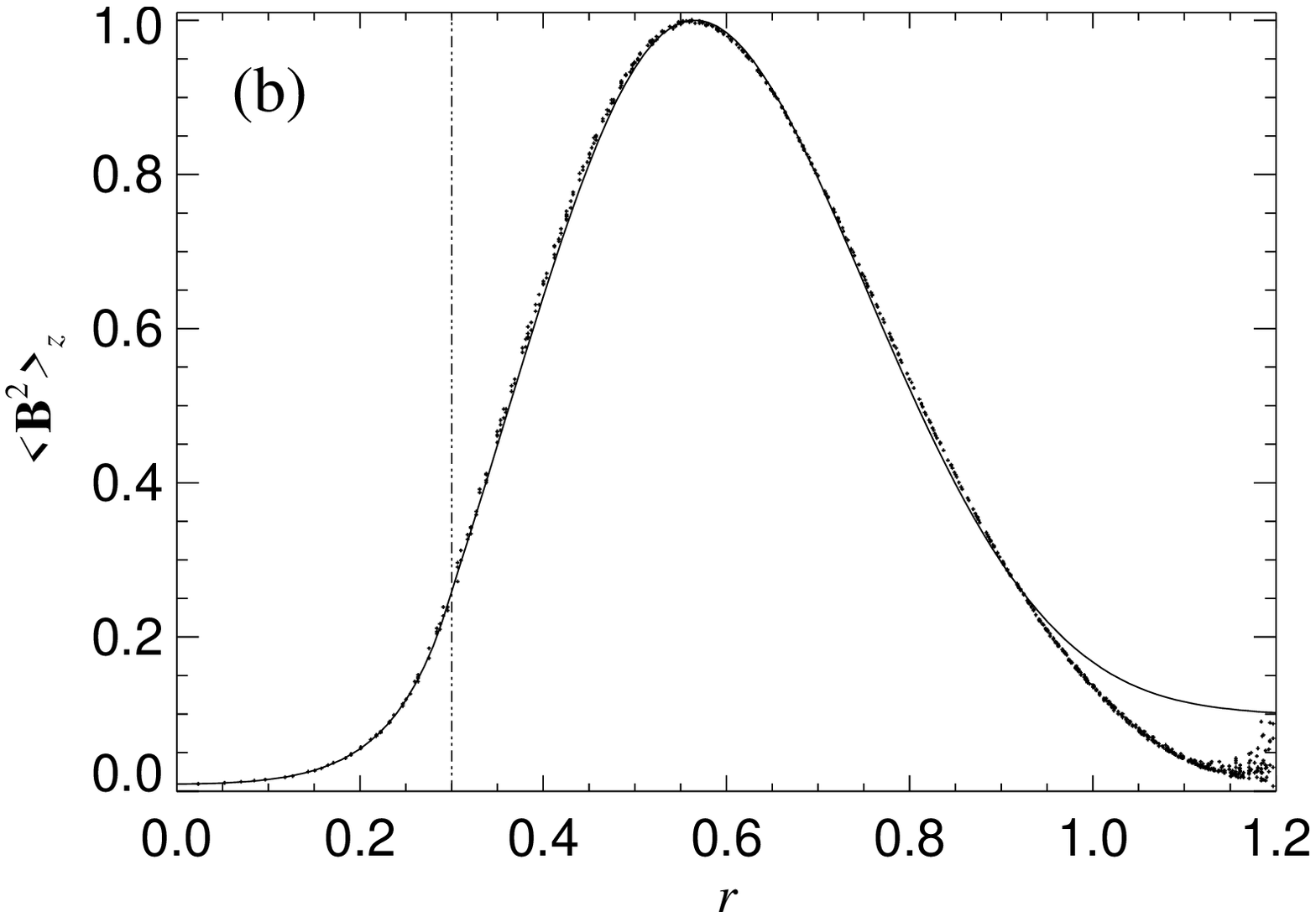}
  \caption{Radial field structure in the kinematic regime
    as obtained from a three-dimensional simulation at a
    resolution
    $\delta x = 0.033$ (dots) and from
    solving the one-dimensional Eq.~(\protect\ref{Pono-1d}) with $R_1 = 0.28$
    (solid).
    (a) Model~1a,
    (b) Model~2a.
    The maximum of $\left<\Bv^2\right>_z$ is normalized to one; radius
    is measured in units of $R_0$.
}
  \label{Fig-B2-r-1d-3d}
\end{figure}

The radial dependence of the (vertically averaged) magnetic energy density
is shown in Fig.~\ref{Fig-B2-r-1d-3d} for the high-resolution runs of
Models~1a and 2a at a time when
the exponential growth has well established itself.
Magnetic energy concentrates in a cylindrical shell of
radius $\RHO_{B}$ and of half-width $\delta\RHO_{B}$, defined
in Eq.~(\ref{Def-r0eff}).
For Model~1a (Fig.~\ref{Fig-B2-r-1d-3d}a), we find
$\RHO_{B} = 0.48$ and $\delta\RHO_{B} = 0.16$ (see also
Table~\ref{Tab-conv-A}).
For comparison, we have overplotted the profile obtained from the
one-dimensional model.

Figure~\ref{Fig-B2-r-1d-3d} shows good agreement between
the three-dimensional and one-dimensional simulations
everywhere except close to the outer boundary.
For $\RHO \approx R_2$, the magnetic energy in the three-dimensional
simulation smoothly turns to zero, while in the one-dimensional calculation
the tangential components of $\Bv$ remain significant for Model~2a.
This is due to the different boundary conditions used ($\Av = \zerov$
vs.~perfectly conducting) and the close agreement of growth rates
and eigenfunctions gives us reason to believe that these
are not influenced by this localized deviation.

\begin{figure}[t!]
  \centering
  \includegraphics[width=0.48\textwidth]{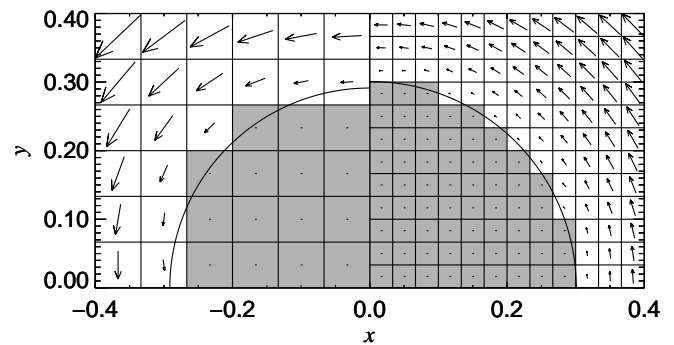}
  \caption{Representation of the inner cylinder (shaded)
on a Cartesian mesh for
    $\delta x = 0.067$ (left) and
    $\delta x = 0.033$ (right).
    Overlayed are the circles of effective radius $R_1^{\rm(eff)}$, and
    arrows representing the velocity field.
    All lengths are measured in units of $R_0$.
    }
  \label{Fig-cyl-disc}
\end{figure}

In Table~\ref{Tab-conv-A}, we compare the eigenvalues $\gammac$ and
the spatial parameters $\RHO_{B}$ and
$\delta\RHO_{B}$ of the magnetic field for
the three-dimensional simulations with those from the
one-dimensional model for different numerical resolutions.
While the high-resolution simulation has only an error of about 5\,\% in
the growth rate $\gammar$, and 0.3\,\% in the frequency
$\gammai$, the lower-resolution run yields errors of 18\,\% and 0.9\,\%,
respectively.
The main source of inaccuracy is the angular
discretization of the inner cylinder boundary as illustrated in
Fig.~\ref{Fig-cyl-disc}.
If we define an effective inner radius $R_1^{\rm(eff)}$ as
the radius of a circle enclosing the same area as
the shaded cells in Fig.~\ref{Fig-cyl-disc}, then
$R_1^{\rm(eff)}\approx0.29$ for the lower-resolution run.
For an inner radius of $0.29$, the one-dimensional model yields a growth
rate that is considerably lower and closer to what is observed in the
lower-resolution run (see Table~\ref{Tab-conv-A}).
This adjustment of $R_1^{\rm(eff)}$ results in better agreement of
$\gammai,\ \RHO_B$ and $\delta\RHO_B$ as well.

\begin{table}[t!]
  \centering
  \newcommand{\mcc}[1]{\multicolumn{1}{c}{#1}}
  \caption{Accuracy of the three-dimensional code, illustrated for
    Model~1a using various resolutions $\delta x$.
    The effective inner radius $R_1^{\rm(eff)}$ is defined in the text.
    Growth rate, frequency, localization radius and width of the magnetic
    field distribution are given for Model~1a at different resolutions
    (top half).
    The lower half of the table shows the accurate values obtained by solving
    the one-dimensional problem (\ref{Pono-1d-rho}), (\ref{Pono-1d-phi})
    for three different values of $R_1$.
    Lengths are measured in units of $R_0$, growth rates and oscillation
    frequencies in units of $\Omega_0$.
  }
  \label{Tab-conv-A}
  \begin{ruledtabular}
  \begin{tabular}{l@{\ecs}lllll}
    \multicolumn{1}{c}{Model}
                           & \mcc{$R_1^{\rm(eff)}$}
                                       & \mcc{$\gammar$}
                                                  & \mcc{$\gammai$}
                                                             & \mcc{$\RHO_{B}$}
                                                                       & \mcc{$\delta\RHO_{B}$}\\
  \colrule
    1a ($\delta x {=}0.067$)
                           & $0.291$   & $0.0144$ & $-1.969$ & $0.467$ & $0.162$ \\
    1a ($\delta x {=}0.033$)
                           & $0.301$   & $0.0160$ & $-1.957$ & $0.478$ & $0.160$ \\
    1a ($\delta x {=}0.017$)
                           & $0.300$   & $0.0162$ & $-1.952$ & $0.483$ & $0.161$ \\
  \colrule
    \strut
    theory ($R_1{=}0.28$)  &           & $0.0140$ & $-1.971$ & $0.468$ & $0.158$ \\
    theory ($R_1{=}0.29$)  &           & $0.0154$ & $-1.961$ & $0.478$ & $0.159$ \\
    theory ($R_1{=}0.30$)  &           & $0.0167$ & $-1.951$ & $0.488$ & $0.160$ \\
  \end{tabular}
  \end{ruledtabular}
  \end{table}

\begin{acknowledgments}
  We are grateful to A.~D.~Gilbert for useful discussions
  and to an anonymous referee for making useful suggestions.
  This work was partially supported by Leverhulme Trust (Grant F/125/AL)
  and PPARC (Grant PPA/G/S/1997/00284). Use of the PPARC supported
  parallel computers in St Andrews and Leicester is
  acknowledged.
\end{acknowledgments}



\bibliography{dynamo1}
\bibliographystyle{apsrev.bst}


\end{document}